%% file: main.tex
\documentclass{article}

\usepackage{arxiv}

\usepackage[utf8]{inputenc} 
\usepackage[T1]{fontenc}    
\usepackage{url}            
\usepackage{booktabs}       
\usepackage{amsfonts}       
\usepackage{bbm}
\usepackage{amssymb}
\usepackage{amsmath}
\usepackage{nicefrac}       
\usepackage{graphicx}
\usepackage[colorlinks]{hyperref}       
\usepackage[nameinlink,noabbrev]{cleveref}
\usepackage{color}
\usepackage{placeins}
\usepackage{doi}
\usepackage[normalem]{ulem}
\usepackage{xspace}
\usepackage{multirow}

\usepackage{caption}
\usepackage{subcaption}
\usepackage{standalone}
\usepackage{tikz}
\usetikzlibrary{shapes,arrows.meta,positioning,calc}

\usepackage{lineno}

\title{Bayesian calibration of interatomic potentials for binary alloys}

\author{
Arun Hegde \\
Sandia National Laboratories\\
Livermore, CA 94551 \\
\texttt{ahegde@sandia.gov} \\
\And
Elan Weiss \\
The Ohio State University\\
Columbus, OH 43210 \\
\texttt{weiss.443@osu.edu} \\
\And
Wolfgang Windl \\
The Ohio State University\\
Columbus, OH 43210 \\
\texttt{windl.1@osu.edu} \\
\And
Habib Najm \\
Sandia National Laboratories\\
Livermore, CA 94551 \\
\texttt{hnnajm@sandia.gov} \\
\And
Cosmin Safta \\
Sandia National Laboratories\\
Livermore, CA 94551 \\
\texttt{csafta@sandia.gov} \\
}

\begin{document}
\maketitle
\begin{abstract}
Developing reliable interatomic potential models with quantified predictive accuracy is crucial for atomistic simulations. Commonly used potentials, such as those constructed through the embedded atom method (EAM), are derived from semi-empirical considerations and contain unknown parameters that must be fitted based on training data. In the present work, we investigate Bayesian calibration as a means of fitting EAM potentials for binary alloys. The Bayesian setting naturally assimilates probabilistic assertions about uncertain quantities. In this way, uncertainties about model parameters and model errors can be updated by conditioning on the training data and then carried through to prediction. We apply these techniques to investigate an EAM potential for a family of gold-copper systems in which the training data correspond to density-functional theory values for lattice parameters, mixing enthalpies, and various elastic constants. Through the use of predictive distributions, we demonstrate the limitations of the potential and highlight the importance of statistical formulations for model error. 

\end{abstract}
\section{Introduction}

In molecular dynamics and atomistic simulation, systems correspond to collections of atoms. The critical quantity of interest is the energy associated with that system, which controls its structure, kinetics, and mechanical properties. There are several methods of varying fidelity for computing this energy. One common strategy is to employ an interatomic potential (IAP). IAPs are functions that take as input the physical coordinates of each atom and return as output the corresponding potential energy of the system. These functions are designed, often heuristically, to induce realistic physical interactions between atoms (such as attraction and repulsion) and contain unknown parameters that must be empirically fitted before the potential can be used. Hence, such models are often highly approximate and the construction of good IAPs has been regarded as something of an art~\cite{brenner2000art}.

The popularity of IAP-based simulation techniques stems from the fact that once properly calibrated, energies and forces between atoms can be computed by simply evaluating the function and its derivatives -- a computationally expedient task. IAPs allow for complex simulations with moderate computational effort, but often at the cost of considerable uncertainty about their validity due to the empirical nature of their construction~\cite{brenner2000art}. In contrast, quantum mechanics-based approaches, such as density-functional theory (DFT), require more burdensome electronic structure calculations and are only practical for systems with relatively few atoms~\cite{goedecker1999linear}. These methods, however, are highly accurate and especially ``transferable'' in the sense that no recalibration is necessary when a specific atom type is embedded in a new structure or joined with other elements. Thus IAPs and DFT constitute opposite ends of the recurring trade-off between tractability and accuracy, where IAP-based methods have become the \textit{de facto} technique for investigating practical-scale material systems by necessity. 

Recently, a number of studies have recognized uncertainty quantification (UQ) as crucial to assessing the reliability of IAP-based simulation, e.g., \cite{frederiksen2004bayesian, angelikopoulos2012bayesian, dutta2018bayesian, longbottom2019uncertainty, patrone2019uncertainty, cailliez2020bayesian, vassaux2021ensembles}. In standard practice, IAP parameters are chosen such that simulation optimally reproduces data for certain preselected quantities of interest (QoIs). This data usually comes from either higher-fidelity theory or experiment and hence may be relatively scarce. Once fitted, the potential is validated by testing how well it predicts other QoIs that were not included during calibration~\cite{cailliez2020bayesian}. Unfortunately, this approach can be lacking due to the single-point nature of the best-fit parameter and associated predictions; there is no consideration of uncertainty in the problem setup and these nominal predictions are not accompanied by confidence intervals or error bars that reflect the actual deviation from the calibration data. A common method for remedying this is by treating unknown quantities, such as model parameters and model errors, as random variables rather than deterministic values. Both calibration and prediction can then be couched in the language of probability theory. In particular, model calibration, \emph{i.e.} parameter inference, can then be carried out using Bayes' rule.

The above issues were first discussed in the specific IAP calibration context by Frederiksen and coauthors~\cite{frederiksen2004bayesian}, who advocated a Bayesian statistical approach to account for the above-mentioned parameter uncertainty, contrasting with commonly used best-fit methods. In particular, they emphasized the importance of quantifying prediction uncertainties to better assess IAP transferability. Since then, a number of studies have explored Bayesian UQ in greater depth for both IAP calibration and prediction. In the context of macromolecular simulation, Cooke and Schmidler~\cite{cooke2008statistical} utilized Bayesian inference to estimate a dielectric force-field parameter. Cailliez and Pernot~\cite{cailliez2011statistical} presented a statistical calibration and prediction framework based on Gaussian error assumptions and then analyzed a simple Lennard-Jones potential for argon. In later work~\cite{pernot2017critical}, the authors examine several different strategies for UQ under the assumption of an inadequate model, using a simple Lennard-Jones potential for krypton as an example. In a two-part study, Rizzi et al.~\cite{rizzi2012uncertainty1,rizzi2012uncertainty2} investigated the impact of parameter uncertainty on both prediction and calibration of a TIP4P model of water with a Lennard-Jones potential. Since atomistic simulation and IAP fitting is a computationally demanding endeavor, which has arguably impeded the broader adoption of UQ in this context, Rizzi et al. proposed polynomial chaos expansions as useful surrogates for the mapping between IAP parameters and macroscale observables as a speed-up to make UQ treatment more feasible. With similar motivation, Angelikopoulos et al.~\cite{angelikopoulos2012bayesian} presented a Bayesian framework specialized to high performance computing environments, exploiting the parallelism afforded by the transitional Markov chain Monte Carlo algorithm. This technique was demonstrated using a Lennard-Jones potential for argon. The studies listed above are a sparse sample of a growing literature -- review articles such as~\cite{cailliez2020bayesian, zhou2017uncertainty, chernatynskiy2013uncertainty} provide broader context to developments in the field.  

The present paper demonstrates how Bayesian UQ can lead to more complete assessments of uncertain model predictions, particularly when the models in question are imperfect, as is often the case with IAPs. Our investigation focuses on potentials of the embedded atom method (EAM) form for binary alloys, i.e., systems composed of two different element types~\cite{ward2012rapid}. We note, however, that there exists a wide variety of interatomic potential models, ranging from extensions of EAM, such as the modified EAM approach~\cite{baskes1987application, baskes1992modified}, to more recent techniques such as SNAP potentials~\cite{thompson2015spectral} and other machine learning based methods~\cite{mueller2020machine}. While our approach and later discussion regarding model inadequacy would be similarly applicable to these other potentials (as all models are approximate), we chose EAM potentials due to their general availability, as evident in online databases such as the Interatomic Potential Repository maintained by NIST~\cite{becker2011atomistic,becker2013considerations,hale2018evaluating} or OpenKIM~\cite{tadmor2011potential, elliott2011knowledgebase}, and their use as building blocks for more complex multi-component potentials~\cite{ward2012rapid}.  Despite their prominent usage, we note that these models often struggle with fitting datasets consisting of multiple physical properties and compositions~\cite{voter1994embedded, grochola2005fitting}. In this context, a principled calibration approach can help endow predictions with an appropriate uncertainty budget to compensate for these limitations.  Thus, while individual parameter evaluations may not match all of the calibration data points, the predictive uncertainty of the calibrated model is such that it can still provide effective coverage.

The novelty of this work is that, to our knowledge, Bayesian calibration methods have hitherto been unexplored in the multi-component alloy setting. As noted in~\cite{cailliez2020bayesian}, current applications of Bayesian techniques have mostly been designed for and tested on simple IAPs, such as Lennard-Jones potentials, and single element systems. Binary alloys, and more generally multi-component systems, offer unique challenges on a variety of fronts including the construction of potentials, the choice of calibration data, and more generally simulation. In our investigation, we encountered two primary obstacles that impeded direct application of standard inference algorithms: the first being the computational cost of individual atomistic simulations, and the second being the lack of strong prior knowledge on appropriate model parameter values. This latter aspect was particularly problematic as it required navigating regions of the parameter space in which the simulation was poorly behaved. We addressed these challenges by implementing an iterative trust-region strategy involving local surrogate models and inference steps. The utility of this scheme was demonstrated through a case study  for Au-Cu alloys in which an IAP model containing $5$ unknown parameters was calibrated with selected DFT-generated physical properties for a variety of compositions. By analyzing the results of this calibration and comparing the associated predictive uncertainty with the data, we expose and discuss key limitations of the model. In addition, this example highlights the role of statistical formulations for model error and their related predictive distributions, which become increasingly important when using an imperfect model. We conclude by going beyond this calibration setting to investigate the model's performance on force predictions, a QoI not included during calibration. Here, we discuss opportunities for model improvement and other future work.

We begin in \Cref{sec:alloy} by reviewing the Rapid Alloy Method for Producing Accurate, General Empirical (RAMPAGE) potentials formalism ~\cite{ward2012rapid} used here to construct binary potentials and detail our simulation setup. Then, \Cref{sec:calibration} casts the calibration problem in a Bayesian statistical setting, with \Cref{sec:strategy} discussing both challenges with the approach as well as our implementation strategy. We conclude with a detailed case study in \Cref{sec:caseStudy} investigating a RAMPAGE potential model for a family of gold-copper alloys, where we show how a ``classical'' set of fitting data \cite{MARTINEZ2013} based on mixing enthalpies, elastic constants, and structural data can be assessed with our approach and compared against force-matching test data. This study highlights how predictive uncertainties enable a more critical and thorough examination of an IAP, both in terms of calibration and prediction.

\section{Alloy design}
\label{sec:alloy}
The promise of discovering new materials with next-generation properties underlies recent interest in multi-component ``high entropy'' or ``concentrated'' alloys~\cite{gao2016high}, where combining similar proportions of elements produces a homogeneous alloy with often excellent mechanical and anti-corrosive properties. These materials are far superior to traditional alloys, such as steel, which are usually constructed by mixing small amounts of solutes with a large quantity of a solvent element. Since good high-entropy alloy systems are rare, the search for them entails exploration of a vast composition space~\cite{cantor2020multicomponent}. The feasibility of these studies, thus, relies on the availability of high quality IAPs to support the required atomistic simulations. We note that even for binary alloys, i.e, systems composed of two element types, the number of readily available potentials with desired element compositions is limited. For example, the NIST Interatomic Potentials Repository~\cite{becker_trautt_hale} provides at least one IAP for all elements of the periodic table up to atomic number 103. Out of the 5253 unique binary systems that these 103 elements could form, binary potentials exist for less than 250 pairs or 5\%. 

In \Cref{sec:eam}, we review the Embedded Atom Method (EAM), a general and popular class of potentials commonly used for binary alloys. The RAMPAGE framework, reviewed in \Cref{sec:rampage}, provides an efficient method for generating multi-component EAM potentials from existing single-element potentials, which are abundant in the literature. Then, \Cref{sec:lammps} details our setup for simulation with these potentials, highlighting the choice of QoIs as well as the workflow.  

\subsection{Embedded Atom Method}
\label{sec:eam}
While many IAP forms exist, this investigation is concerned with Finnis-Sinclair Embedded Atom Method (EAM) potentials \cite{voter1994embedded,finnis1984simple}. EAM potentials are widely used in molecular dynamics simulation because they provide a convenient platform for balancing both modeling flexibility and computational efficiency. As we are dealing with binary alloys, we focus on the situation with two different element types, denoted by $A$ and $B$ in the expressions below.

Within the EAM framework, the potential energy of an atom $i$ of species $\alpha \in \{A,B\}$ is given by, 
\begin{equation}
    \label{eq:EAM}
    E_i =     \frac{1}{2}\sum_{j \neq i} V_{\alpha \beta} \left( r_{ij} \right) + F_{\alpha}\left(\sum_{j \neq i}\rho_{\alpha \beta}\left(r_{ij}\right)\right),
\end{equation}
where $\beta \in \{A,B\}$ is the species associated with atom $j$. For clarity, this labeling convention is typically shorthand for $\alpha = \alpha({i})$ and $\beta = \beta({j}) $. In the above, $V_{\alpha\beta}$, $\rho_{\alpha\beta}$, and $F_{\alpha}$ are the pair potential, electron density, and embedding functions. It follows that the total potential energy of a system is given by $E_{\rm tot} = \sum_{i}E_i$. 

EAM enhances simple pair potentials, which consist just of the first summation on the right hand side of \Cref{eq:EAM}, by adding a ``many-body'' embedding term $F_{\alpha}$ that takes as input the local ``electron density'' $\rho = \sum_{j \neq i} \rho_{\alpha \beta} (r_{ij})$.  $F_\alpha$ is interpreted as the energy required to place or embed an atom of species $\alpha$ into the existing electron density due to the other atoms in the system. As shown, this density $\rho$ sums the contributions from all atoms within the cutoff radius, which are ``many'' and hence EAM is sometimes referred to as a many-body potential. 

Note that in the standard EAM framework developed by Daw and Baskes~\cite{daw1984embedded}, the electron density of a binary $AB$ system is described by two functions $\rho_{A}$ and $\rho_{B}$, which specify the contributions of elements $A$ and $B$ to the local electron density. The Finnis-Sinclair family of EAM potentials, which we have described above, extends upon this by introducing species specific local electron density contributions where $\rho_{\alpha\beta}$ is the contribution of atom $j$ to the electron density around atom $i$ and depends on the element of both atoms~\cite{finnis1984simple}. 

This caters to a more modern interpretation in which the EAM concept is categorized as a bond-order potential where bond strength changes with the number of bonds formed by an atom. Here, the role of the density function is to count the effective number of neighbors; nearby neighbors contribute more than distant neighbors, and bonds get weaker as more bonds are formed. Introducing a dependence on the second element into the density function allows for an additional degree of flexibility -- namely, the effect that a neighboring atom has on a central atom's bond strength now depends on the element types of both atoms. For a binary system, a Finnis-Sinclair EAM potential therefore contains nine functions: three pair potentials $V_{A A}$, $V_{B B}$, and $V_{A B}$, two embedding functions $F_{A}$ and $F_{B}$, and four electron density functions $\rho_{A A}$, $\rho_{B B}$, $\rho_{A B}$, and $\rho_{B A}$. 

The nine component functions can each be specified by any functional form, ranging from simplified analytical expressions to lookup tables and spline functions, as long as they are univariate and combine in the manner described by \Cref{eq:EAM}. While this gives the modeler significant freedom in developing expressions to accurately model alloys, computational efficiency is preserved by ensuring that the resulting IAP depends only on pairwise distances, as is the case with traditional pair potentials. This is in contrast to more complex IAPs that directly incorporate many-body terms, such as a dependence on bond angles between three atoms. These attributes give EAM-type potentials a unique advantage over most other techniques, though under the constraint that pair potential formulations strongly favor close-packed crystal structures. Hence, the use cases for EAMs mostly include metals with face-centered cubic (FCC) and sometimes hexagonal close-packed (HCP) structures \cite{Agrawal_2013,Agrawal_2015}, although it has been shown that BCC metals can be stabilized by the inclusion of a large number of neighbor shells within the cutoff radius~\cite{jin2015}.

\subsection{RAMPAGE potentials for binary alloys}
\label{sec:rampage}
The Rapid Alloy Method for Producing Accurate General Empirical Potentials (RAMPAGE Potentials) is a computationally efficient IAP generation system designed to produce binary and multi-component EAM potentials from preexisting single element potentials~\cite{ward2012rapid}. By starting from published elemental potentials, the scope of fitting a binary EAM IAP is reduced to the three cross species functions, $V_{A B}$, $\rho_{A B}$, and $\rho_{B A}$. The fitting problem can be further reduced by selecting analytical functions to model the three cross interactions. Details of the models selected for this work are presented in \Cref{sec:caseStudy}. 

The efficiency of the RAMPAGE system \cite{Roth2017} stems from the fact that single element IAPs are readily available through online databases such as the NIST Interatomic Potential Repository~\cite{becker2011atomistic,becker2013considerations,hale2018evaluating} or OpenKIM~\cite{tadmor2011potential, elliott2011knowledgebase}. Hence, new models can be quickly generated by changing the single element tables or the functional forms. While using published elemental IAPs drastically reduces the overhead in generating new binary IAPs, there are specific challenges involved in ensuring compatibility of the chosen elemental tables. In order to be combined and used by MD packages such as the Large-scale Atomic/Molecular Massively Parallel Simulator (LAMMPS)~\cite{plimpton1995fast}, the elemental tables must have identical spacing and cutoff. In addition to the requirements imposed by MD packages, the potential energy ranges of the elemental tables selected must match to produce physically meaningful cross interactions. In EAM potentials, the density functions $\rho_{A}$ and $\rho_{B}$, lie on an arbitrary unit scale and are used only as inputs for the embedding functions $F_{A}$ and $F_{B}$. Furthermore, there are no requirements on the form of the single element pair potentials $V_{A}$ and $V_{B}$; for example, they may range from being repulsive at all distances to being attractive everywhere.

Using Voter's invariant transformations, the form of these functions can be manipulated while preserving the overall behavior of the elemental IAPs~\cite{voter1986accurate}. To ensure compatibility of the density functions, the following Voter invariant transformations are applied,
\begin{equation}
\label{eq:rhotransform}
\begin{aligned}
     & \rho_i \left( r \right) \xrightarrow{} s \cdot \rho_i \left(r \right)\\
     & F_i \left( \rho \right) \xrightarrow{} F_i \left( \frac{\rho_i}{s} \right). 
\end{aligned}
\end{equation}
The value $s$ is selected to transform the domain of $F_i$ to a common maximum value, which here is chosen as 400. This particular value was chosen to ensure good numerical performance while normalizing the density and embedding functions of the two elemental IAPs to the same unit scale \cite{riegnerThesis}. The pair potentials are likewise transformed by
\begin{equation}
\label{eq:phitransform}
\begin{aligned}
     & F_i \left( \rho \right) \xrightarrow{} F_i \left( \rho_i \right) + g \cdot \rho_i \left( r \right) \\
     & V_i \left( r \right) \xrightarrow{} V_i(r) - 2g \cdot \rho_i \left( r \right),
\end{aligned}
\end{equation}
where $g$ is chosen here to set the final pair potential minimum to 0.04\% of the cohesive energy. From there, the elemental IAPs are interpolated after splining to ensure matching grid spacing and cutoff. When combining potentials of different cutoff radii, The final cutoff radius is equal to the largest elemental cutoff radius.

A Finnis-Sincliar type binary IAP can now be constructed from these standardized elemental IAPs by inserting the cross interaction terms $V_{A B}$, $\rho_{A B}$, and $\rho_{BA}$. Within the RAMPAGE workflow, these are modeled using analytical functions with free parameters, which we assemble into a vector $\theta$. Binary IAPs can then be used in MD simulations as described in \Cref{sec:lammps}. Since there are no three-body terms, the binary potentials can then be further combined into multi-component potentials, for example, to handle concentrated alloys~\cite{ward2012rapid,weiss2022RAMPAGE}, for which pairwise partitioning has been shown to be valid from DFT calculations \cite{OBERDORFER2019}.

The training set in this investigation is composed of DFT computed properties for 17 face centered cubic (FCC) structures for an Au-Cu system spanning the composition range between the two elements. We chose this system because it is fully miscible over the entire composition range and at the same time shows ordering effects at lower temperatures. To best approximate ideal solid solution alloys in the limited simulation sizes available to DFT, 32 atom special quasi-random structures (SQS)~\cite{von2010generation} were employed for the training set. For each SQS the mixing enthalpy, lattice constant, bulk modulus, and elastic tensor components are computed using the Vienna \textit{Ab-initio} Simulation Package (VASP)~\cite{kresse1994ab, kresse1996efficiency, kresse1996efficient, kresse1999ultrasoft} as described in \Cref{sec:appendix_dft}.

Similarly to the functions of the elemental IAPs, the training set for the intermediary compositions must also be standardized to ensure that the fitting problem is physically well-defined. Aside from the invariant transformations applied above, the elemental IAP functions are unmodified from their original published forms and treated as fixed inputs to the model. It follows that properties of the elemental endpoints are therefore  unaffected by the model parameters and constant throughout the calibration process. Thus, to remove any initial and unalterable discrepancy, the DFT data at the elemental endpoints for lattice constant, bulk modulus, and elastic tensor components are normalized to the  corresponding elemental IAP values. This rescaling is then propagated to the the QoIs for the intermediate compositions by a linear rule of mixtures (akin to Vegard's law), such that for given normalization values $c_{A}={\rm QoI}_A^{\rm IAP}/{\rm QoI}_A^{\rm DFT}$ and $c_{B}={\rm QoI}_B^{\rm IAP}/{\rm QoI}_B^{\rm DFT}$, which are determined by the pure elements, the intermediate values of $c$ are given by
\begin{equation}
\label{eq:ruleofmixtures}
    c_{x} = (1-x)\,c_A + x\,c_B,
\end{equation}
where $x$ is the atom fraction of B atoms in a given SQS with composition A$_{1-x}$B$_x$. For the mixing enthalpies, the cohesive energies are calculated from the DFT values by subtracting the energy of the atoms in their elemental state from the total energy of the SQS cell. IAPs by default calculate cohesive energies, and so scaling for the mixing enthalpies happens through scaling according to the cohesive energies of the elemental endpoints. 

\subsection{Simulation workflow}
\label{sec:lammps}

For a given configuration of atoms, an IAP model is used to compute the corresponding potential energy of the system. As such, an IAP defines the potential energy surface over the space of atomic configurations. Different choices of parameter values lead to different realizations of this surface, thus inducing different physical behaviors and properties of the resulting system. In this context, calibration amounts to identifying model parameters which produce systems whose physical properties match a set of DFT-generated reference values.

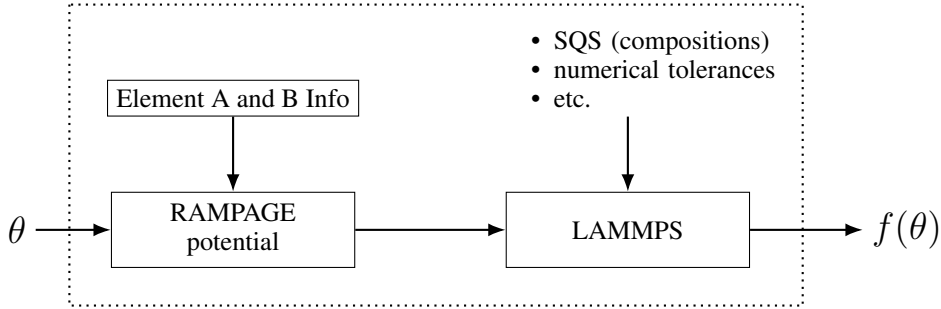
\begin{figure}[hbtp]
\centering
\input{simulation_workflow.tex}
\caption{Simulation workflow for a generic binary system. Here, $\theta$ represents the input parameter vector for the examined IAP, and $f(\theta)$ the vector of output QoI values.
}
\label{fig:sim_workflow}
\end{figure}

In the present work, the QoIs are taken at the equilibrium structure, i.e., the configuration of atoms with minimal potential energy. Specifically, we are interested in the physical properties of lattice parameter, mixing enthalpy, bulk modulus, and the elastic constants $C_{11}$, $C_{12}$, $C_{44}$, 

which we compute using LAMMPS~\cite{plimpton1995fast}. The simulation workflow, mapping from an input vector of model parameters to the output physical properties, is illustrated below in \Cref{fig:sim_workflow}. 
\FloatBarrier
For a given parameter vector $\theta$, the RAMPAGE potential is assembled and fed into LAMMPS. As highlighted earlier, RAMPAGE potentials are intended for use over the entire sensible composition range and hence, for a fully miscible system as examined here, calibration should span the range from a pure $A$ to a pure $B$ system. This amounts to computing the above physical properties for a representative set of compositions, which are each individually modeled using fixed, deterministic, 6912-atom SQSs. Here, each structure consists of a 12$\times$12$\times$12 supercell of a conventional FCC unit cell and contains the appropriate number of $A$ and $B$ atoms corresponding to the desired composition. These SQS cells are fully relaxed in LAMMPS by conjugate gradient minimization with the IAP for the given parameter vector $\theta$. The lattice parameter and mixing enthalpy QoIs are computed directly from the resulting groundstate structure. Elastic constants -- bulk modulus, $C_{11}$, $C_{12}$, $C_{44}$ are then calculated as described in Appendix~\ref{sec:appendix_dft}. In the following sections, we refer to this entire sequence of steps, from $\theta$ to the output QoIs $f(\theta)$, as the LAMMPS simulation. Additional notation is introduced in \Cref{sec:calibration} to index the various QoIs.

\section{Bayesian calibration}
\label{sec:calibration}

Bayesian calibration is an effective paradigm for handling UQ in scientific and computational models~\cite{kennedy2001bayesian,higdon2004combining, higdon2008computer, bayarri2007framework, sargsyan2015statistical}. As highlighted in the introduction and references therein, traditional methods for calibration in atomistic simulation focus on finding best fit parameter values using criteria such as least squares. In contrast, Bayesian inference approaches \cite{bernardo2009bayesian, gelman1995bayesian} provide a more general framework for calibration in which parameter and model uncertainties can be accounted for in a principled manner. Rather than seeking out a single optimal parameter, Bayesian methods start with the specification of a prior probability density over the parameter space -- representing the uncertainty in the parameter values -- and a likelihood function -- which can be used to rank parameter values based on their agreement with the observed data. The combination of a prior and likelihood constitutes a probabilistic model. Incorporating both likelihood and prior through Bayes' formula leads to a posterior probability density on the parameter space, which represents the updated uncertainty in the parameters now having accounted for the observed data. On the computation side, the primary objective in a Bayesian calibration procedure is to generate and analyze samples from this posterior distribution.   

Here, we discuss the application of Bayesian methods to the problem setup described in the previous section. In \Cref{sec:formulation}, we formulate a statistical model for the LAMMPS simulation in which the error between the simulation and DFT data is modeled as a Gaussian random variable. This formulation allows a simple characterization of the likelihood, and therefore the posterior distribution over the parameters. \Cref{sec:strategy} then identifies the simulation runtime coupled with the behavior of QoI responses as key challenges in our setup that combine to obstruct direct application of standard sampling algorithms. Indeed, issues with respect to runtime could be alleviated with some loss in accuracy, e.g., by using smaller cells (we currently use $12$x$12$x$12$) or reducing the convergence tolerances for energy minimization. Moreover, developing IAPs that optimally balance this cost-accuracy tradeoff is an ongoing area of research   \cite{plimpton2012computational,zuo2020performance}. \Cref{sec:strategy} further details how these difficulties motivate a multi-stage strategy involving locally-fit surrogate models and inference steps.  

\subsection{Formulation}
\label{sec:formulation}
We model the relation between the (normalized) DFT data $y_i(x)$ and the LAMMPS simulation $f_i(x; \theta)$ as follows,
\begin{equation}
\label{eq:probmodel}
\begin{aligned}
     & y_{i}(x) = f_{i} (x; \theta) + \epsilon_i(x)  \\
     & \epsilon_i(x) \sim 
     \begin{cases}
     \mathcal{N}\left(0, \sigma^2 f_{i} (x; \theta)^2\right) & \textrm{for } i \in
     \mathcal{I}_1 := \{\textrm{lattice constant},\textrm{C}_{11},\textrm{C}_{12},\textrm{C}_{44},\textrm{bulk modulus}\} \\
      \mathcal{N}\left(0, \sigma^2 f_{i} (x; \theta)^2 + \tau^2 \right) & \textrm{for }  i \in \mathcal{I}_2 := \{\textrm{mixing enthalpy}\} 
     \end{cases} \\
     & x \in \mathcal{X}, 
\end{aligned}
\end{equation}
where $\mathcal{X}$ denotes the discrete set of compositions, $\theta$ is the vector of model parameters, and the subscript $i$ ranges over the recorded physical properties. Taken together, the combination $(i,x)$ indexes the calibration QoIs. $\mathcal{N}(\cdot,\cdot)$ denotes the normal distribution with the specified mean and (co)variance. The terms $\epsilon_i(x)$ provide a statistical description of the discrepancy or ``model error'' between the DFT data and LAMMPS outputs and are, in the present work, modeled as independent Gaussian random variables whose standard deviations are proportional in magnitude to the simulation output. The additional parameter $\sigma$ describes this level of proportionality. We make this modeling choice to account for the different units and magnitudes among the QoIs. Additionally, the auxiliary variance term involving $\tau$ is included as a regularization since the mixing enthalpy can take on small values that straddle zero and which may otherwise lead to degeneracy in the corresponding $\epsilon_i(x)$. We note that in the context of our present formulation, we also consider the model error to account for possible errors or imperfections in the DFT data.

The statistical formulation in \Cref{eq:probmodel}, which we sometimes refer to as the \emph{data model} to distinguish it from the \emph{physical model} $f$, defines the likelihood as the following multivariate normal,
\begin{equation}
\label{eq:likelihood}
\begin{aligned}
p(\textbf{y} | \theta, \textrm{ln} \sigma, \textrm{ln} \tau ) &= \mathcal{N} \left(\textbf{f}(\theta), C \right),
\end{aligned}
\end{equation}
in which
\begin{equation}
\label{eq:likelihood_cov}
\begin{aligned}
& C_{\mathcal{I}_1} = \textrm{diag}\left( \{\sigma^2 f_i(x; \theta)^2\}_{x \in \mathcal{X}, i \in \mathcal{I}_1} \right) \\
& C_{\mathcal{I}_2} = \textrm{diag}\left( \{\sigma^2 f_i(x; \theta)^2 + \tau^2\}_{x \in \mathcal{X}, i \in \mathcal{I}_2} \right) \\
& C = \begin{bmatrix} C_{\mathcal{I}_1} & 0 \\ 0 & C_{\mathcal{I}_2} \end{bmatrix}.
\end{aligned}
\end{equation}
Note that in the above, $\textbf{y}$ and $\textbf{f}(\theta)$ are used as shorthand for the vector with entries corresponding to the different $(i,x)$ combinations. The ordering is implied by the context. For convenience, in subsequent expressions we group the parameters together in the extended vector $\lambda = \{ \theta, u, v\}$, where $u=\ln \sigma$ and $v=\ln \tau$. In this way, the standard deviations $\sigma = \exp{u}$ and $\tau = \exp{v}$ are ensured to always be positive. We refer to the model error parameters $\ln \sigma$ and $\ln \tau$ as \textit{hyper}parameters in our setup, as is common in the literature, e.g. \cite{bishop2006pattern,mackay1996hyperparameters,sargsyan2015statistical}, to highlight that they are external to the physical model $f$ and separate from the model parameters $\theta$. 

Inference in this probabilistic setting requires a fully specified probabilistic model, which is a joint distribution describing the relationships among all variables of interest. In this case, these variables are the DFT data $\textbf{y}$ and the parameters $\lambda$. Ideally this distribution, which we denote with the probability density function $p(\textbf{y}, \lambda)$, conveys all of the uncertainties inherent in the problem, including those pertaining to the data, parameters, and model structure. A convenient way of building such a model is by factoring $p(\textbf{y}, \lambda)$ into the product of a likelihood (or conditional density) $p(\textbf{y} | \lambda)$, which we defined earlier for our problem in \Cref{eq:likelihood}, and a prior density $p(\lambda)$:  $p(\textbf{y}, \lambda) = p(\textbf{y} | \lambda) p(\lambda)$. Note that prior $p(\lambda)$ allows one to specify the \textit{a priori} uncertainty in the parameter values. We discuss our prior choice further in \Cref{sec:strategy}.  

The setting for model calibration is that we have observed specific values for $\mathbf{y}$ and are therefore interested in the conditional distribution on the parameters $\lambda$ given this information, i.e, $p(\lambda | \mathbf{y})$. This is known as the posterior distribution and can be computed via Bayes formula,
\begin{equation}
\label{eq:bayes}
\begin{aligned}
p(\lambda | \textbf{y}) &= \frac{p(\textbf{y}, \lambda)}{p(\textbf{y})} \\
&=\frac{p(\textbf{y} | \lambda)}{p(\textbf{y})} p(\lambda).
\end{aligned}
\end{equation}
The posterior distribution describes our updated or \textit{a posteriori} uncertainty in the parameters after having accounted for the data $y$ (in contrast with the \textit{a priori} uncertainty). As shown in \Cref{eq:bayes}, this transformation from prior to posterior is accomplished by weighting the prior with a term describing the influence of the observed data. The normalization factor $p(\mathbf{y})$ is a proportionality constant and ensures that the posterior density integrates to one, 
\begin{equation}
\label{eq:evidence}
p(\textbf{y}) = \int  p(\textbf{y} | \lambda)p(\lambda) d\lambda.
\end{equation}
The quantity $p(\textbf{y})$ is also referred to as the marginal likelihood, or Bayesian evidence, and can be used for model selection \cite{kass1995bayes,mackay1992bayesian}. Hence if we were comparing different IAP models, then the explicit computation of the integral in \Cref{eq:evidence} would become important. 

Once calibrated, a model can then be to make predictions. For example, diagnostics such as model checking ~\cite{gelman1995bayesian,gelman1996posterior} can be performed by computing the posterior predictive distribution (PPD) over prospective outputs $\tilde{y}$,
\begin{equation}
\label{eq:ppd}
\begin{aligned}
p_{\textrm{ppd}}(\tilde{y}| \textbf{y} ) &= \int  p(\tilde{y} |  \lambda, \textbf{y} )p(\lambda| \textbf{y} ) d\lambda 
=\int  p(\tilde{y} |  \lambda)p(\lambda| \textbf{y} ) d\lambda,
\end{aligned}
\end{equation}
where we have used that $\tilde{y}$ and $\textbf{y}$ are conditionally independent given parameters $\lambda$. The PPD can be sampled by simply pushing posterior samples from $p(\lambda| \textbf{y} )$ through the statistical model in \Cref{eq:probmodel}.

Alternatively, predictions can be made using purely the physical model $\mathbf{f}(\theta)$ by evaluating samples from marginal posterior for the model parameters $p(\theta|\textbf{y})$. The resulting distribution is called the push forward posterior (PFP) ~\cite{sargsyan2015statistical} and is given by,
\begin{equation}
\label{eq:pfp}
p_{\textrm{pfp}}(\tilde{y}| \textbf{y}) = \int  \delta\left( \tilde{y} - \textbf{f}(\theta) \right) p(\theta | \textbf{y}) d\theta,
\end{equation}
where $\delta(\cdot)$ is the delta function.  

The previous predictive distributions were all with respect to the calibration QoIs. Given a new prediction QoI $q = F(\theta)$ -- for example, representing some physical property whose model $F$ relies on the same set of model parameters $\theta$ -- the corresponding PFP is,
\begin{equation}
\label{eq:pfp2}
p_{\textrm{pfp}}(\tilde{q}| \textbf{y}) =
\int  \delta\left( \tilde{q} - F(\theta) \right) p(\theta | \textbf{y}) d\theta.
\end{equation}

\subsection{Inference strategy}
\label{sec:strategy}

A calibrated model is fully characterized by its posterior distribution $p(\lambda| \textbf{y})$. Therefore, the general inference target is to obtain samples from this posterior, which can then also be used for tasks such as prediction. In most practical cases, such analysis requires Markov chain Monte Carlo (MCMC) algorithms~\cite{liu2008monte}. Given data $\textbf{y}$, a typical Bayesian inference approach involves specifying a prior and likelihood, and then performing MCMC to collect a sufficient number of posterior samples for good characterization of the posterior density. In our application, a direct implementation of this strategy presents some challenges.

First, MCMC algorithms often require $\mathcal{O}(10^5-10^6)$ steps to reach satisfactory convergence. Each step involves evaluating the likelihood, which can quickly become a bottleneck if the forward model is expensive to compute. For example, an evaluation of our LAMMPS simulation, i.e., computing all $102$ QoIs for a given parameter vector, can take up to several hours on a single CPU. While this burden may be addressed to some extent by relaxing convergence tolerances and decreasing the cell size in the simulation (at the cost of a potential loss in accuracy), reductions in computation time may not be possible with more complex IAPs  \cite{plimpton2012computational}. Hence, directly including the exact likelihood in \Cref{eq:likelihood} is generally not a feasible strategy. A common workaround, which we exploit, is to replace the underlying simulation $f$ with a cheaper-to-evaluate surrogate $g$ constructed through regression~\cite{sacks1989design,kennedy2001bayesian}. In this way, the surrogate can be trained offline using a moderate number of simulation runs, and then used in lieu of the simulation during MCMC. Of course, much like reducing the model complexity, the savings do not come for free -- the accuracy of the resulting posterior density fundamentally depends on the quality of the surrogate fit~\cite{stuart2018posterior}.

Second, we often don't have enough knowledge to specify a well-motivated prior distribution at the outset. For example, while we may have nominal parameter values associated with the pure elemental potentials, how these values relate to the cross species terms which constitute our model is often unclear. In such cases, it is common to use an uninformed prior over a wide range. The challenge here is that the parameter ranges for which the simulation is ``well-behaved'' or even amenable to surrogate modeling are also often unknown. This can lead to difficulties if sampling from the posterior requires evaluating regions in which the simulation behaves erratically. In the present work, we have observed that large portions of the parameter space can lead to poor results, e.g., unrealistic equilibrium structures and QoI values with negligible likelihood. Building surrogate models that are globally accurate over a wide prior range is both challenging and inefficient. What is required is a more focused approach that adaptively targets regions of high posterior probability.  

\begin{figure}[hbtp]
     \centering
     \begin{subfigure}[b]{0.33\textwidth}
         \centering
         \includegraphics[width=\textwidth]{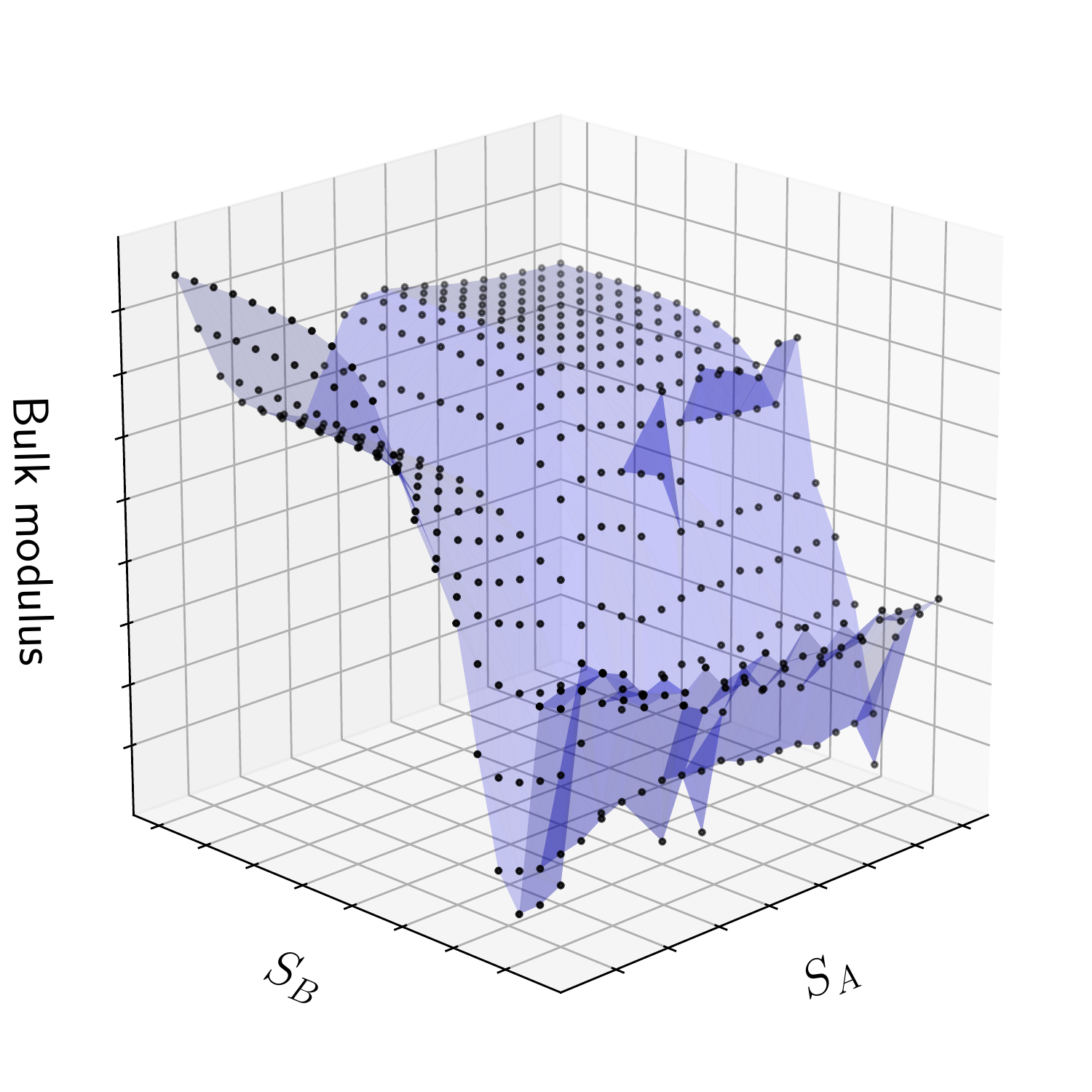}
         \caption{}
         \label{fig:slice1}
     \end{subfigure}
     \hfill
     \begin{subfigure}[b]{0.33\textwidth}
         \centering
         \includegraphics[width=\textwidth]{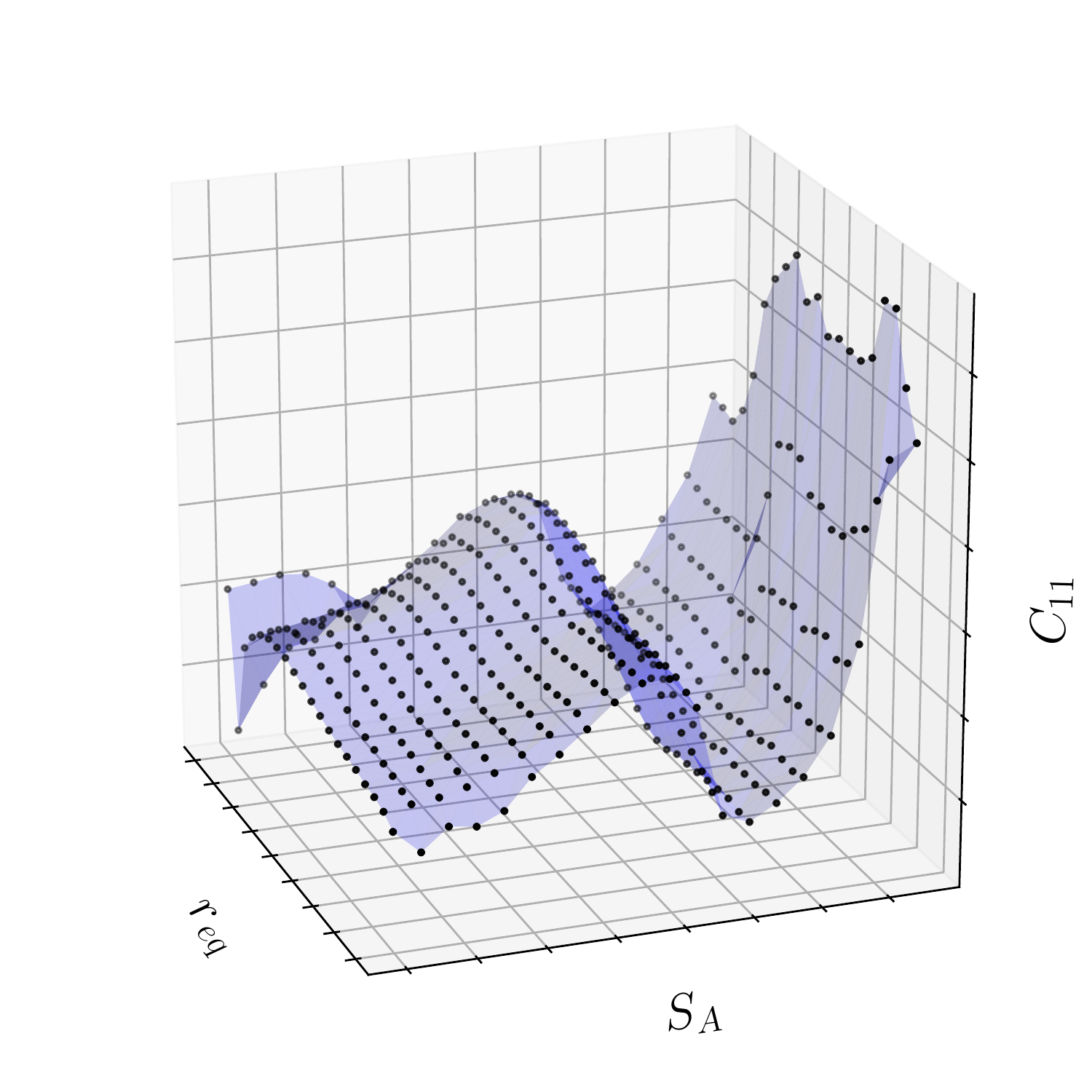}
         \caption{}
         \label{fig:slice2}
     \end{subfigure}
     \hfill
     \begin{subfigure}[b]{0.33\textwidth}
         \centering
         \includegraphics[width=\textwidth]{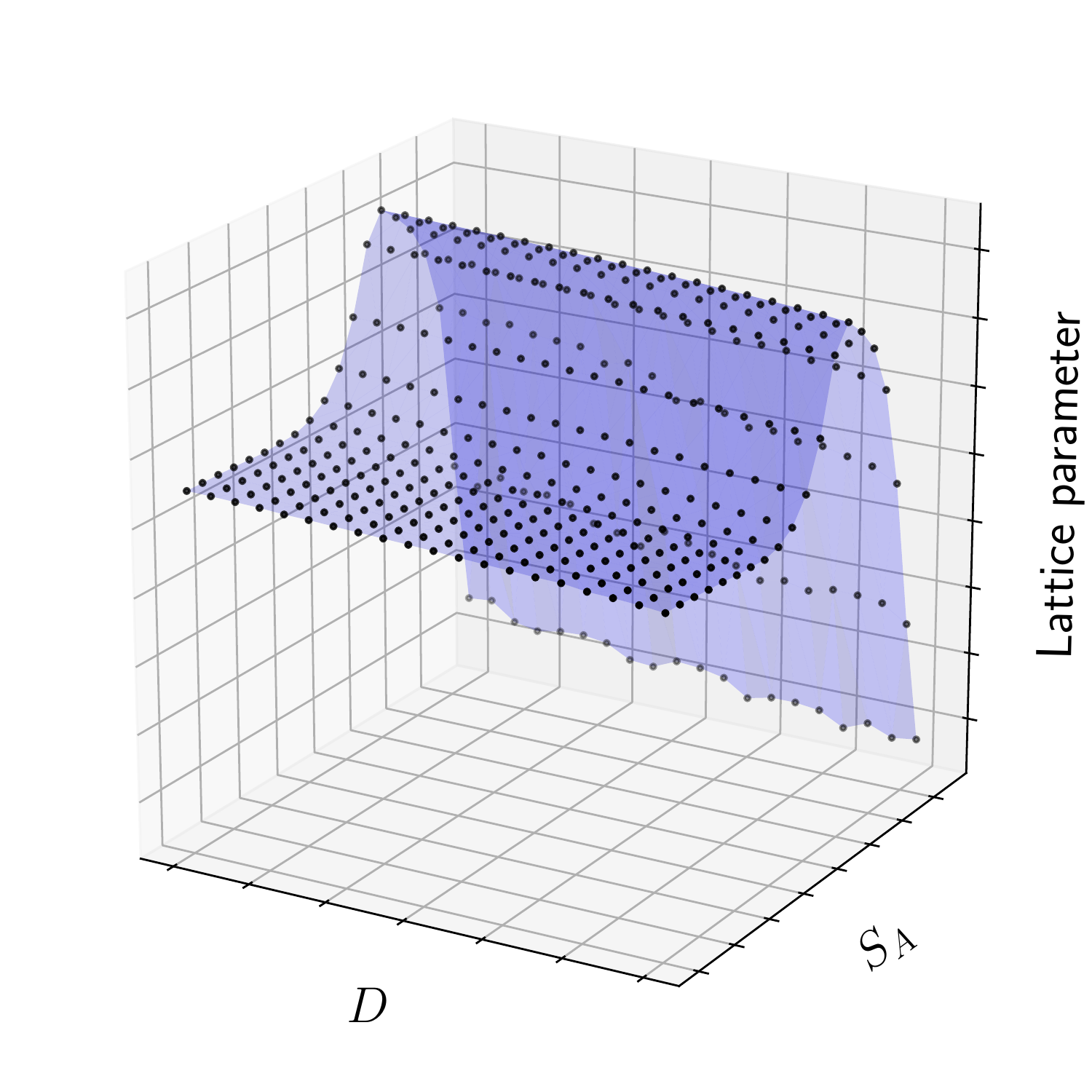}
         \caption{}
         \label{fig:slice3}
     \end{subfigure}
        \caption{Illustration of responses for different QoIs at a fixed composition ($A_{0.5}B_{0.5}$) across two-dimensional slices of the parameter space.}
        \label{fig:2Dresponses}
\end{figure}

In \Cref{fig:2Dresponses}, below, we illustrate slices of various QoI responses taken from the $5$-parameter model described later in \Cref{sec:caseStudy}. These slices were generated by fixing a parameter vector in what we believed would be a reasonable region (which, we note, contains the posterior discussed in \Cref{sec:caseStudy_calibration}), varying pairs of parameters, and then evaluating the indicated QoIs in LAMMPS. These figures display several notable characteristics -- e.g., presence of outliers and non-smoothness (\Cref{fig:slice1}), flat regions followed by steep regions (\Cref{fig:slice1,fig:slice2,fig:slice3}), etc. -- all of which can lead to difficulties when constructing surrogate models.   
\FloatBarrier
How can a ``good'' parameter region, i.e., one of high posterior probability, be found? As indicated above, a one-shot approach to inference over a wide prior region is not feasible. Instead, we adopt the following strategy based on a sequence of locally fit surrogates: 
\begin{enumerate}
    \item Initialize a box $\mathcal{H}$ in the model parameter space. This may be chosen, for example, as centered around a ``best fit'' point or derived from domain expert opinion. 
    \item Initialize a collection of samples $\{\theta^m\}_{m=1}^M$ in $\mathcal{H}$, e.g., via Latin hypercube sampling \cite{mckay1979comparison}.
    \item Evaluate the sample set in LAMMPS to generate the corresponding QoIs $\{(\theta^m, \textbf{f}^m)\}_{m=1}^M$.
    \item Remove any outliers and other ``bad'' parameter vectors from the sample set. A simple criterion would be to prune parameters whose corresponding QoI evaluations deviate significantly from the DFT data. To avoid extra indexing notation, we assume in the following that no samples need to be removed.
    \item Fit surrogate models $g_i$ such that $g_i (x;\theta^m) \approx f_i(x; \theta^m)$ for all QoIs $(x,i)$. This typically involves splitting the sample set into training points, for fitting, and test points, for assessing fitting errors. If the surrogate models have unacceptably large errors, then shrink the box dimensions and repeat. 
    \item Draw samples from the posterior distribution, e.g., using an MCMC algorithm. Importantly, the likelihood is approximated by using the surrogate models in lieu of the the LAMMPS simulation. The prior is set to be uniform and to contain $\mathcal{H}$. Such specification indicates that we have no preference for any particular value in the prior -- i.e., a priori all included parameter values are weighted equally.  
    \item Extend/translate/shrink the box $\mathcal{H}$ so as to contain the posterior samples.
    \item Append the posterior samples (or a subset of them) to the data set and remove samples that are outside the updated $\mathcal{H}$.
    \item Repeat steps (3)-(8) until:
    \begin{itemize}
        \item The box $\mathcal{H}$ does not change between consecutive iterations (the posterior appears completely contained).
        \item The surrogate modeling error is low.
    \end{itemize}
\end{enumerate}
By continually refocusing the box and appending new ``posterior'' samples, the accuracy of the surrogate models improves in regions of high posterior probability. We note that if any of the above steps indicate the posterior distribution has well-separated modes, then maintaining a single box becomes ineffective. In such cases, the above strategy would need to spawn and maintain multiple boxes. Throughout the above procedure, LAMMPS evaluations are performed in batches, which is conducive to runs on computing clusters with queuing systems. Finally, we note that the above steps are used as a template. For example, one may choose to only fit surrogate models for certain QoIs or to include new Latin hypercube samples in the updated box during the iterations. Moreover, the choice of which sampling algorithm to use in step $6$ depends on characteristics observed during the analysis.

\begin{figure}[hbtp]
     \centering
     \begin{subfigure}[b]{0.3\textwidth}
         \centering
         \includegraphics[width=\textwidth]{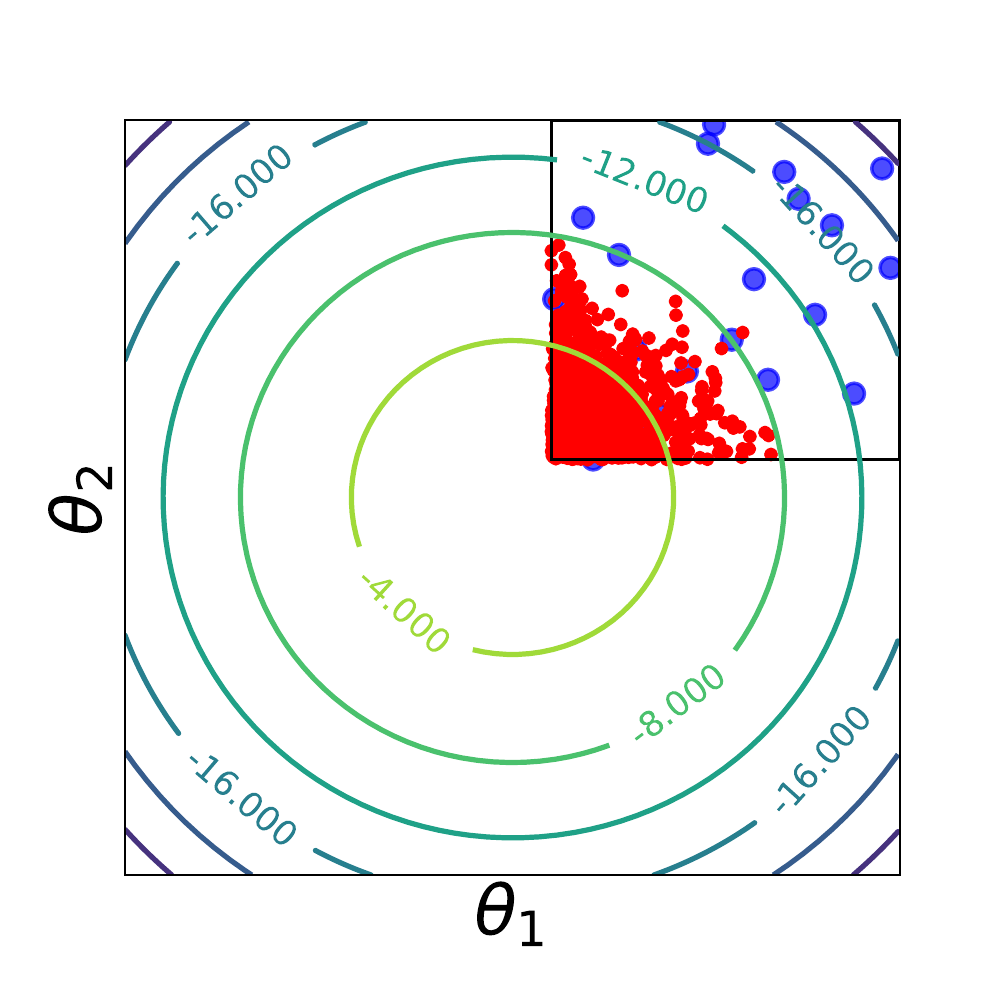}
         \caption{}
         \label{fig:toyEx1}
     \end{subfigure}
     \hfill
     \begin{subfigure}[b]{0.3\textwidth}
         \centering
         \includegraphics[width=\textwidth]{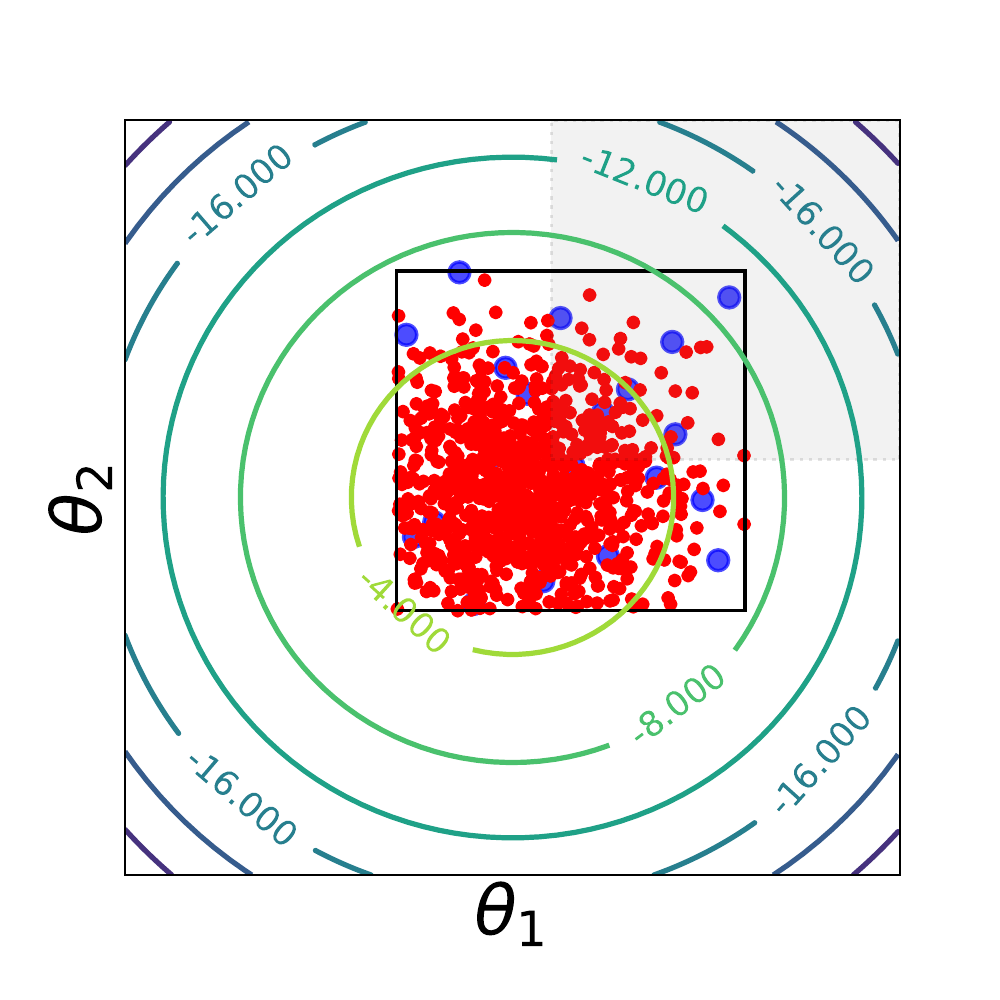}
         \caption{}
        \label{fig:toyEx2}
     \end{subfigure}
     \hfill
     \begin{subfigure}[b]{0.3\textwidth}
         \centering
         \includegraphics[width=\textwidth]{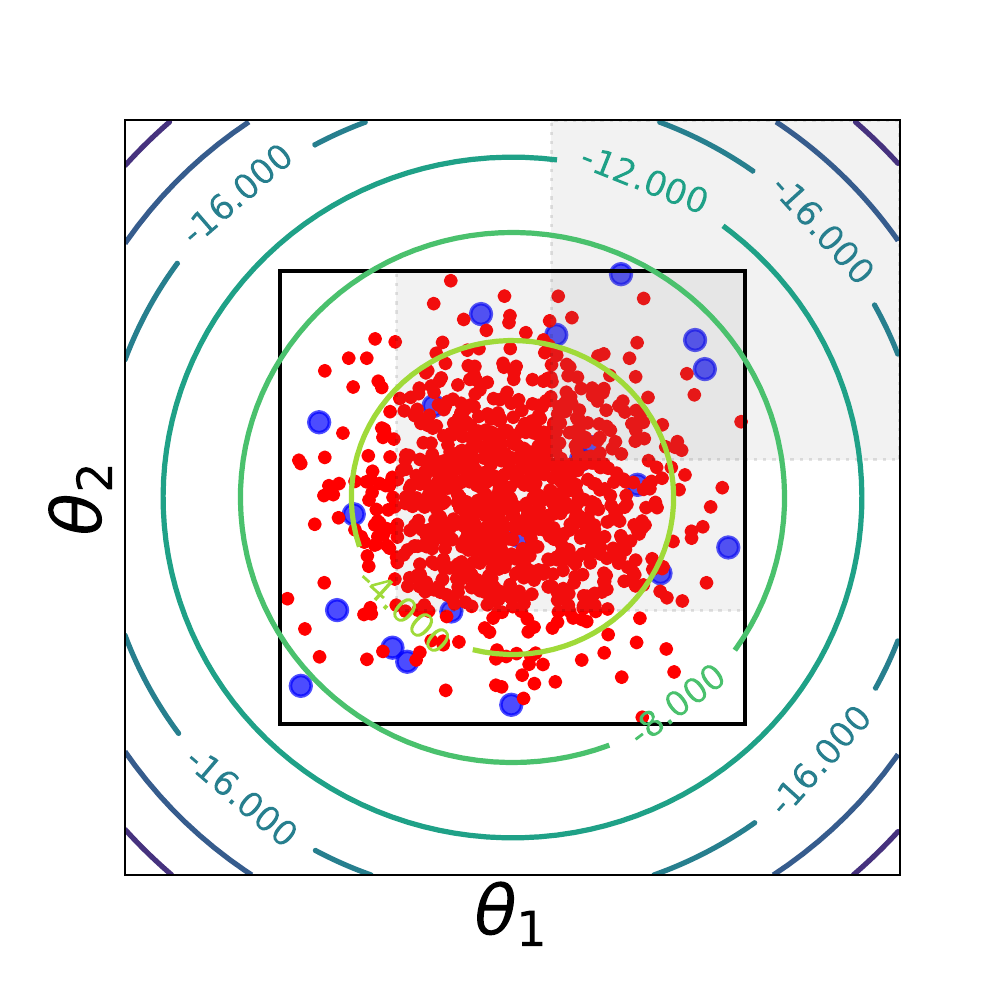}
         \caption{}
        \label{fig:toyEx3}
     \end{subfigure}
    \caption{Toy example illustrating the movement of $\mathcal{H}$ during the inference strategy.}
     \label{fig:toyEx}
\end{figure}

\Cref{fig:toyEx} provides a provides a simple illustration of the above procedure, focusing on the movement of the box $\mathcal{H}$. In this toy example, the statistical model is given by $y = \theta + \epsilon$ where $\theta \in \mathbb{R}^2$, $\epsilon \sim \mathcal{N}(0,I)$, and with $y= [0, 0]^\intercal$ as the observed value. Note that although our general procedure utilizes surrogate models and surrogate-based approximations to the likelihood, in this example we can directly implement the exact likelihood. The region $\mathcal{H}$ is initialized in \Cref{fig:toyEx1} in the upper right corner. The blue points correspond to Latin hypercube samples and would typically be used to construct the surrogate models (as in steps $3$-$5$ above). The red points are samples from the resulting posterior distribution when a uniform prior is employed over $\mathcal{H}$. Finally, the contours illustrate various level sets of the log likelihood function. Following the procedure, $\mathcal{H}$ is adjusted based on the posterior samples, which in this case corresponds to translating and enlarging the original $\mathcal{H}$, leading to the results in \Cref{fig:toyEx2,fig:toyEx3}. The end result is a region $\mathcal{H}$ that more aptly contains the posterior samples. In realistic examples, however, the transition from \Cref{fig:toyEx1} to \Cref{fig:toyEx3} might require several iterations of the procedure to ensure well-fitting surrogates and proper coverage of the MCMC samples.   
\FloatBarrier
\section{Case study: Au-Cu binary alloys}
\label{sec:caseStudy} 

 To demonstrate the method, we applied the above strategy for calibration to investigate a RAMPAGE potential model for a family of gold-copper (Au-Cu) binary alloys. As discussed previously in \Cref{sec:rampage}, we chose this particular system as it is miscible across the entire composition range and also displays ordering effects in the low temperature regime. Following the RAMPAGE formalism, the alloy potential was constructed from a pair of existing elemental Au and Cu EAM potentials developed by Zhou \emph{et al.}~\cite{zhou2004misfit}. In their original work, these elemental potentials were fitted to material properties such as lattice constants, elastic constants, bulk moduli, vacancy formation energies, and sublimation energies and are therefore compatible with our current setup. The modellable components of our EAM alloy potential thus correspond to the two inter-element electron density functions and one pair potential, which we chose as  
\begin{equation}
\label{eq:voter}
\begin{aligned}
& \rho_{BA}(r) = r^6 \left( \exp(-S_A r) + 2^9 \exp(-2 S_A r) \right)\\
& \rho_{AB}(r) = r^6 \left( \exp(-S_B r) + 2^9 \exp(-2 S_B r) \right),
\end{aligned}
\end{equation}
and
\begin{equation}
\label{eq:morse}
V_{AB}(r) = D \left( \exp \left( -2 \alpha \left( r - r_{\text{eq}} \right) \right) - 2 \exp \left( - \alpha \left( r - r_{\text{eq}} \right) \right) \right).
\end{equation}
Electron densities of this form have been used in prior work by Voter~\cite{voter1993alamos}, and the pair potential is a standard Morse potential. Following Voter \cite{voter1993alamos}, the pair potential was then modified to ensure smooth behavior at the cutoff $r_{cut}$, 
\begin{equation}
\label{eq:cutoff}
V_{AB}^{\textrm{smooth}}(r)=V_{AB}(r)-V_{AB}(r_{cut}) + \left(\frac{r_{cut}}{m}\right) \left[ 1 - \left( \frac{r}{r_{cut}} \right)^m \right] \left( \frac{dV_{AB}}{dr} \right)_{r=r_{cut}}.
\end{equation}
To comply with previous notation, the model parameters are collected in the vector $\theta = \begin{bmatrix} r_{eq}, D, \alpha, S_A, S_B \end{bmatrix}^\intercal$. 

The Au-Cu system and the above IAP serve as a useful platform to investigate concepts discussed in the Introduction, particularly, the utility of predictive uncertainty when working with an approximate model. In this context, our focus here is not on developing a better potential, but rather endowing an existing one with reliable predictive uncertainties that accurately capture the discrepancy from training data. To that end, our case study serves as a demonstration. We show in \Cref{sec:caseStudy_calibration} that Bayesian calibration with the above IAP can provide meaningful predictive distributions, despite also showing that the physical model is deficient in several aspects. In doing so, we highlight the crucial role played by the model error term of \Cref{eq:probmodel}. Moreover, in \Cref{sec:caseStudy_forces}, we discuss challenges and future directions for extending this beyond the calibration setting. We briefly overview our results in the following two paragraphs.

In \Cref{sec:caseStudy_calibration}, we present Bayesian calibration results for this potential. Recall that in \Cref{sec:strategy}, the inference is conducted by building and using surrogate models rather than the LAMMPS code directly. Note that in this study, we use Gaussian process surrogates \cite{rasmussen2006gaussian, sacks1989design, kennedy2001bayesian}. We discuss these results first, highlighting the use of surrogate-based predictive distributions as diagnostics for the calibration. The outcome of the calibration is a posterior probability distribution on the parameters $\lambda$, which quantifies their uncertainty. Having calibrated our potential, we then propagate this parameter uncertainty to a corresponding set of posterior distributions for the QoIs. By focusing on the predictive quality of these distributions and the range of discrepancies from the DFT data, we highlight the role of model error or inadequacy in the IAP. Specifically, we demonstrate that the predictive performance is highly variable and depends on both composition and physical property, reflecting the fact that the model fit is being traded off among the QoIs. Moreover, the pushforward posterior distributions -- which rely on just the physical model and neglect model error -- are shown to be highly over confident. In contrast, the posterior predictive distributions -- which include model error -- provide a more appropriate level of uncertainty that captures the actual degree of discrepancy with the calibration data. Hence, model error is required for reliable prediction of the calibration QoIs. The overall results in this section highlight the limitations of the IAP formulated above and clarify the utility of statistical constructions for model error~\cite{kennedy2001bayesian, brynjarsdottir2014learning, sargsyan2015statistical, sargsyan2019embedded}.

In \Cref{sec:caseStudy_forces}, we explore the performance of the calibrated model in predicting interatomic forces in order to highlight limitations of the IAP, some of which were initially exposed during calibration. Indeed, the calibration exercise summarized above and further described in the next section has two primary findings. First, the predictions of the calibrated physical model -- as exemplified by the pushforward posteriors -- display varying ranges of discrepancy when compared with the DFT data. Second, differences between the pushforward posterior and posterior predictive distributions, as will be discussed in \Cref{sec:caseStudy_calibration}, demonstrate that the appropriate level of uncertainty is not communicated by just the marginal posterior on the model parameters $p(\theta|\textbf{y})$. Model error is required to have predictive uncertainties that better reflect the data. This, coupled with the facts that (a) force data was not included during calibration and (b) the additive model error in \Cref{eq:probmodel} is tailored to the calibration and hence does not extend to forces, makes reliable prediction of these interatomic forces a challenge. We further discuss these results in the light of predictive uncertainties, highlighting the need for a more flexible IAP structure and, additionally, embedded forms of model error which can more readily transfer to predictions beyond the calibration setting~\cite{sargsyan2015statistical, sargsyan2019embedded}.

\subsection{Calibrating an IAP}
\label{sec:caseStudy_calibration}
 
\begin{figure}[hbtp]
    \centering
    \includegraphics[width=0.8\textwidth]{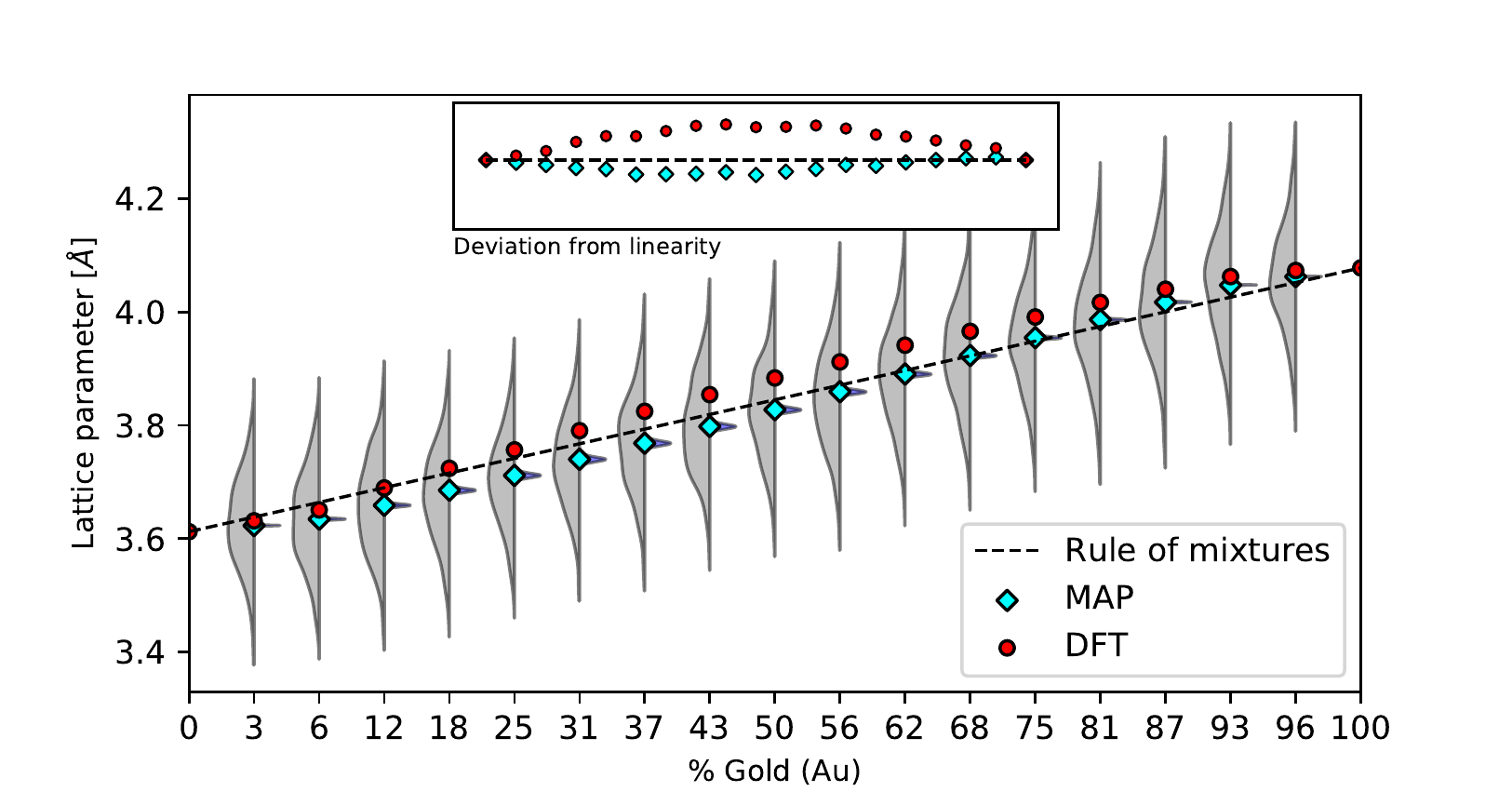}
    \caption{Posterior predictive (left, gray) and pushforward posterior (right, blue) distributions using the surrogate models for the lattice parameter.}
    \label{fig:surrogate_latticePar}
\end{figure}

\begin{figure}[hbtp]
    \centering
    \includegraphics[width=0.8\textwidth]{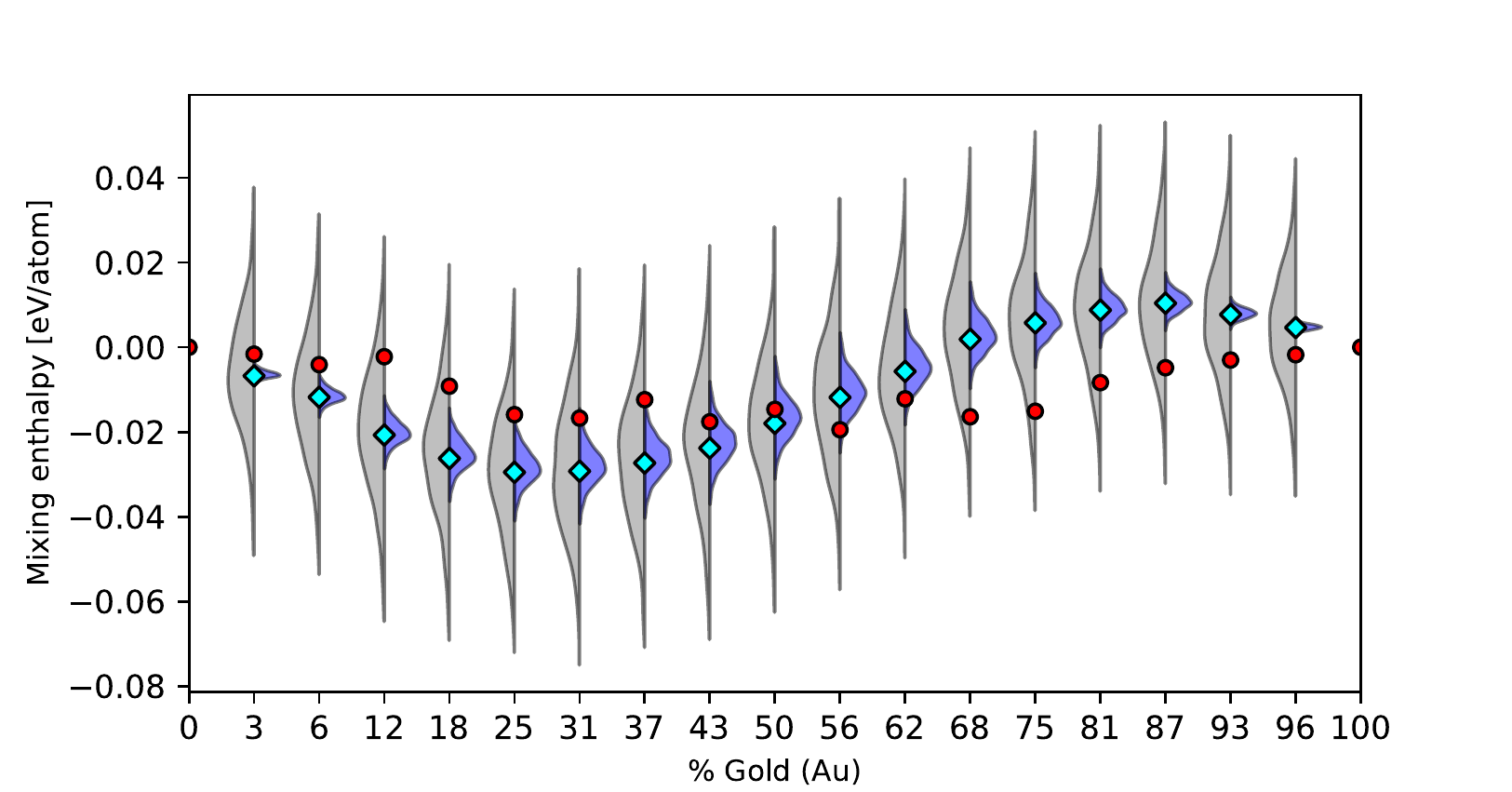}
    \caption{Posterior predictive (left, gray) and pushforward posterior (right, blue) distributions using the surrogate models for the mixing enthalpy.}
    \label{fig:surrogate_mixEnthalpy}
\end{figure}

\begin{figure}[hbtp]
    \centering
    \includegraphics[width=0.8\textwidth]{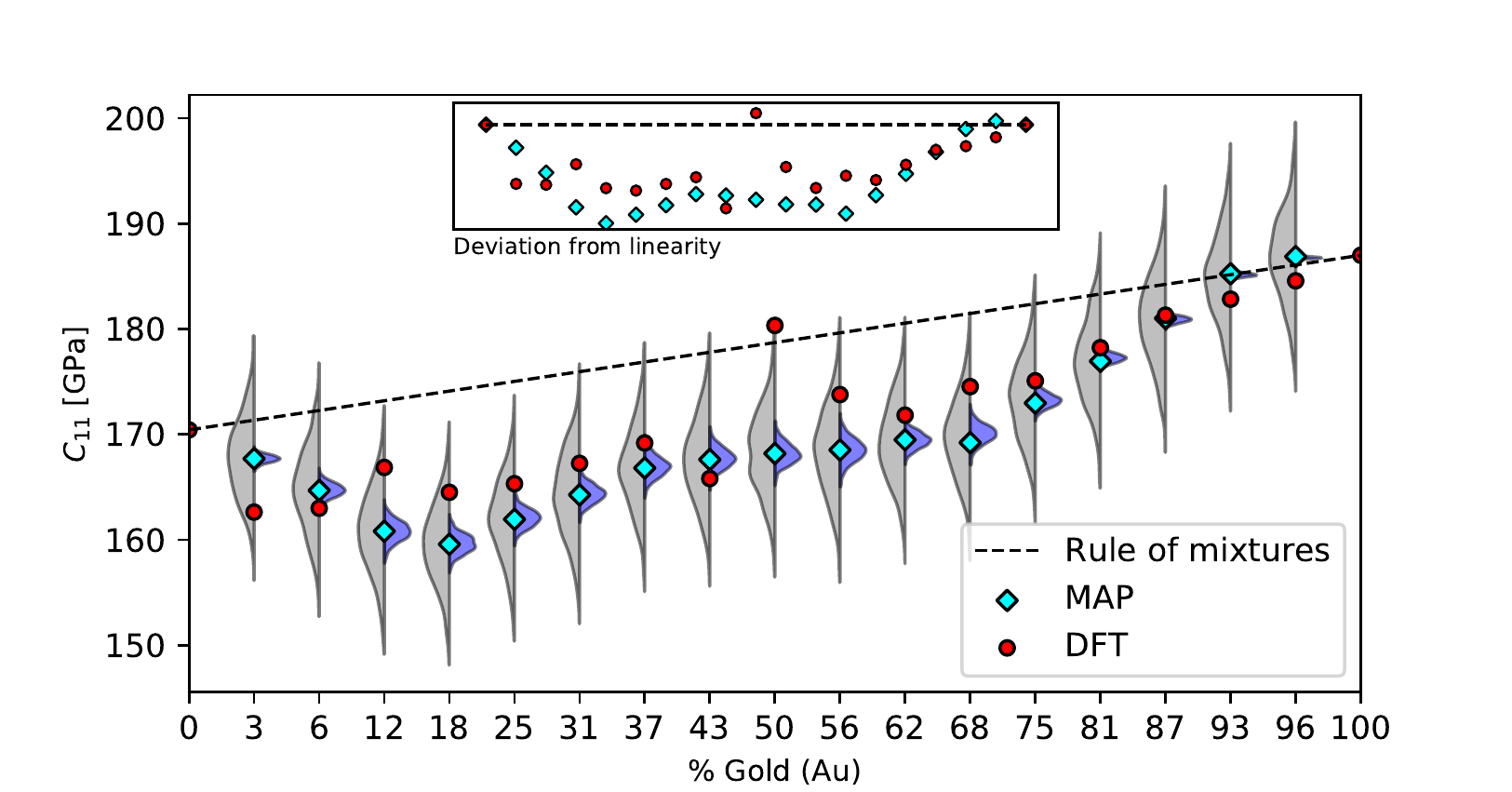}
    \caption{Posterior predictive (left, gray) and pushforward posterior (right, blue) distributions using the surrogate models for $C_{11}$.}
    \label{fig:surrogate_C11}
\end{figure}

\begin{figure}[hbtp]
    \centering
    \includegraphics[width=0.8\textwidth]{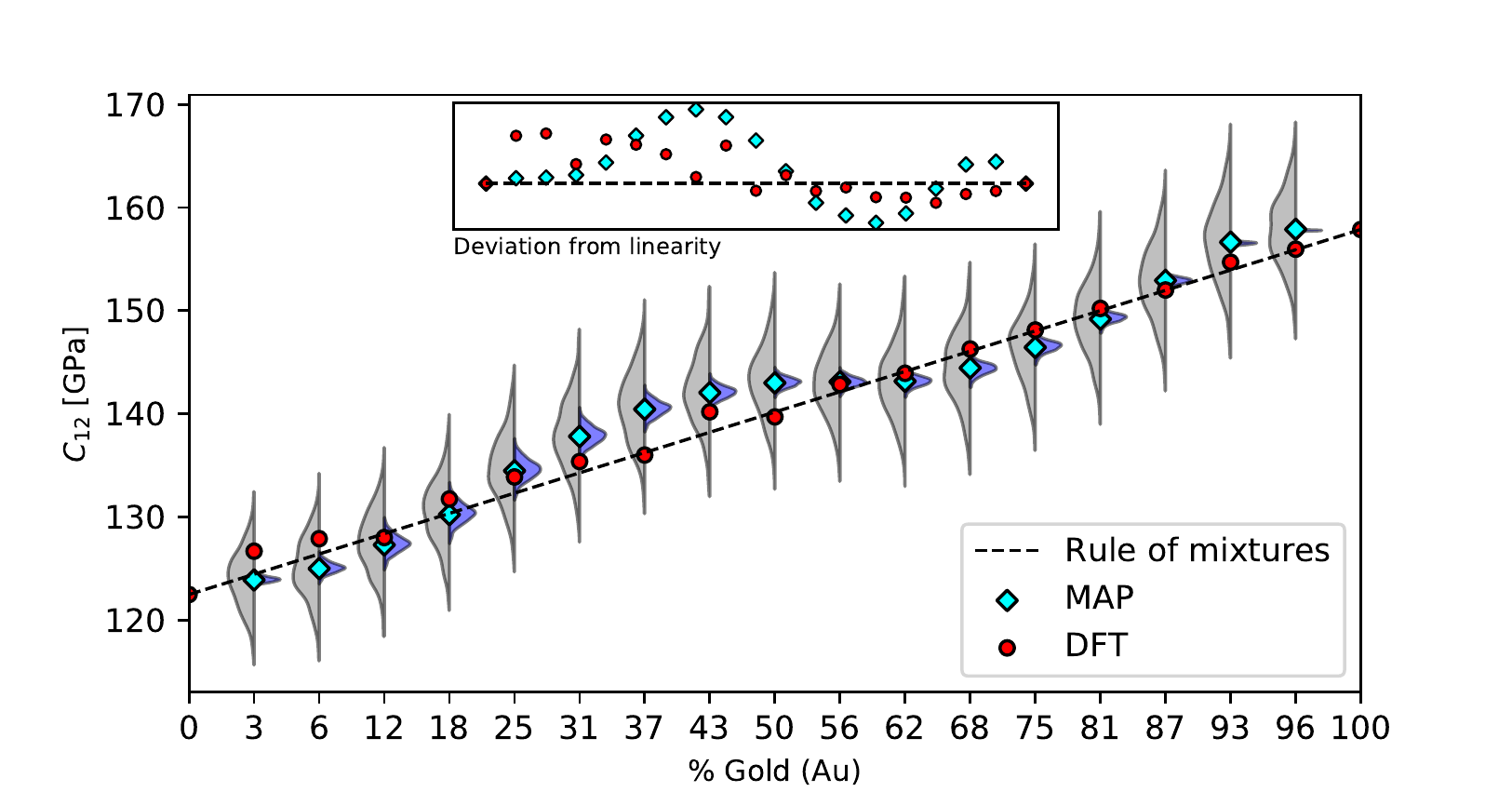}
    \caption{Posterior predictive (left, gray) and pushforward posterior (right, blue) distributions using the surrogate models for $C_{12}$.}
    \label{fig:surrogate_C12}
\end{figure}

\begin{figure}[hbtp]
    \centering
    \includegraphics[width=0.8\textwidth]{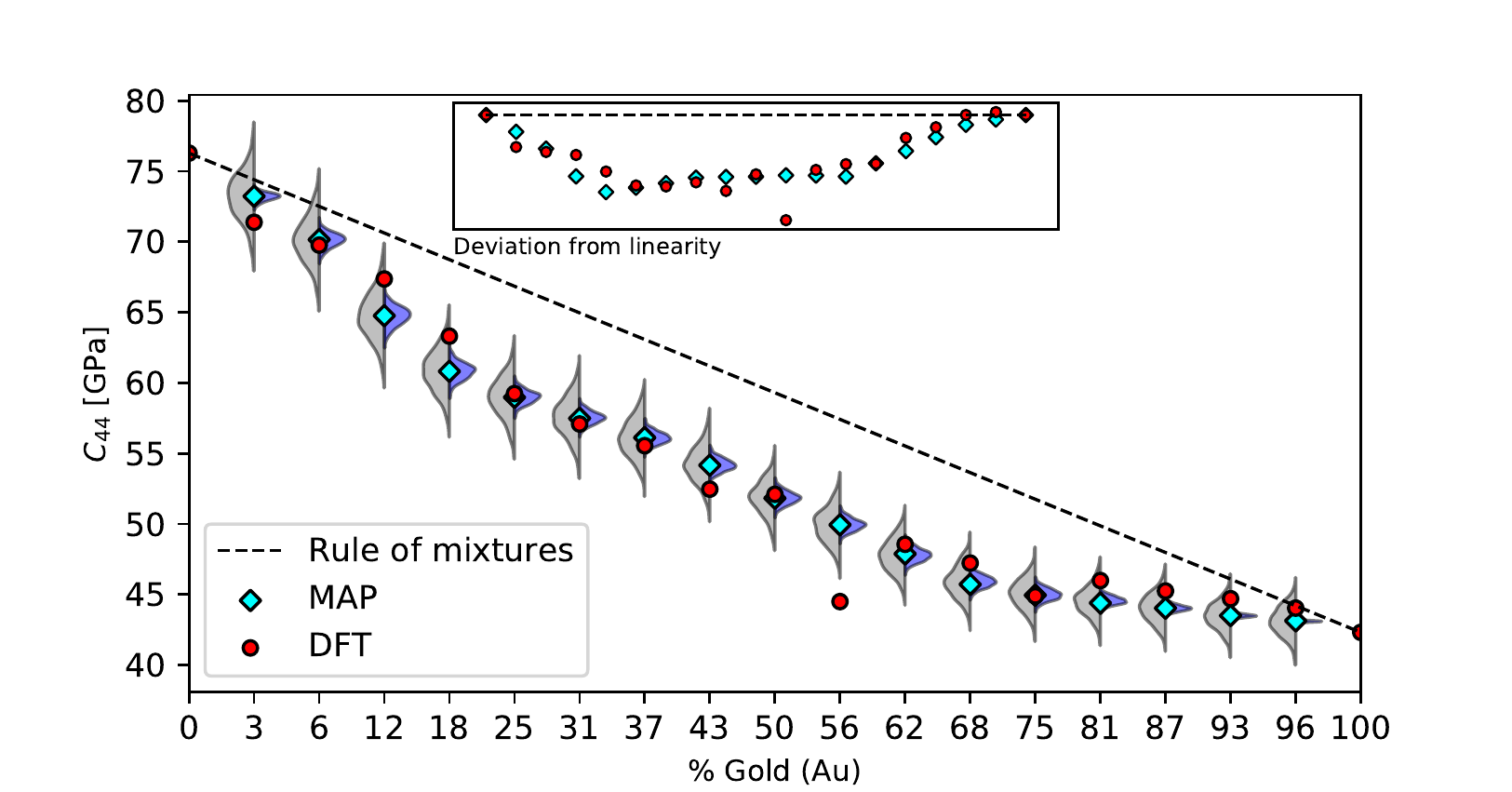}
    \caption{Posterior predictive (left, gray) and pushforward posterior (right, blue) distributions using the surrogate models for $C_{44}$.}
    \label{fig:surrogate_C44}
\end{figure}

\begin{figure}[hbtp]
    \centering
    \includegraphics[width=0.8\textwidth]{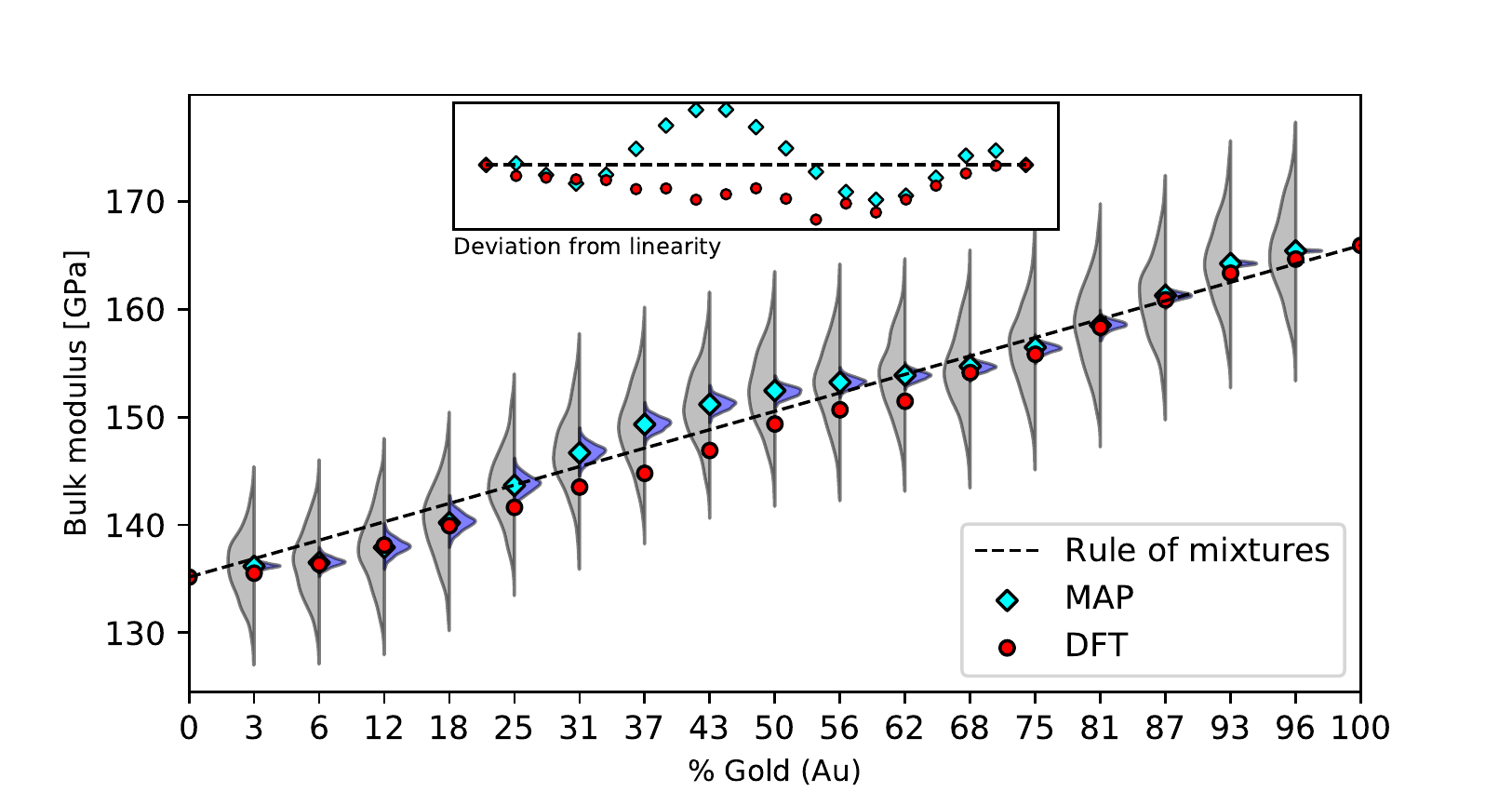}
    \caption{Posterior predictive (left, gray) and pushforward posterior (right, blue) distributions using the surrogate models for the bulk modulus.}
    \label{fig:surrogate_bulkMod}
\end{figure}

The calibration QoIs were selected at the outset of the investigation to span $17$ different compositions $x \in \{\frac{1}{32}, \frac{2}{32}, \frac{4}{32}, \hdots, \frac{28}{32}, \frac{30}{32},\frac{31}{32}\}$,
ranging from $3\%$ to $97\%$ gold, with $6$ physical properties per composition. Hence, there were $102$ QoIs in total. As discussed in \Cref{sec:lammps}, the physical properties corresponded to lattice parameters, mixing enthalpies, elastic constants ${C}_{11}$, ${C}_{12}$, ${C}_{44}$, and bulk moduli. This choice of QoIs is fairly ``classical'' and similar selections have been used since the early years of IAP fitting \cite{voter1986accurate,voter1994embedded}. The calibration data consisted of a corresponding set of DFT computed values, one for each of the $102$ QoIs. Recall that as a preprocessing step, the DFT data required a corrective scaling in order to be comparable with LAMMPS outputs. The normalization was discussed in \Cref{sec:rampage} and is to be interpreted as part of the model. Our case study also highlights the fact that the DFT model, which is used to generate the calibration data, is itself imperfect due to the various theoretical and numerical approximations that go in to its formulation. These imperfections appear in the calibration data, shown later in \Cref{fig:surrogate_latticePar,fig:surrogate_mixEnthalpy,fig:surrogate_C11,fig:surrogate_C12,fig:surrogate_C44,fig:surrogate_bulkMod}, as non-smooth composition-dependent fluctuations for certain physical properties. Similar characteristics have been observed in DFT simulation of alloy properties, for example in work by Wong \emph{et al.} \cite{wong2018optimizing} examining the accuracy of using SQSs of varying size. These imperfections in both the model and corresponding data provide
additional motivation for the statistical construction of model error in \Cref{eq:probmodel}. In the present work, we handle the error associated with both the IAP model and the DFT data in a joint fashion. Following the procedure in \Cref{sec:strategy}, the calibration was conducted using surrogate models in lieu of the LAMMPS simulation. Each stage of the analysis, however, required LAMMPS evaluations to generate training (and test) data for the surrogates.

\begin{figure}[hbtp]
    \centering
    \includegraphics[clip,trim=0 0 0 0]{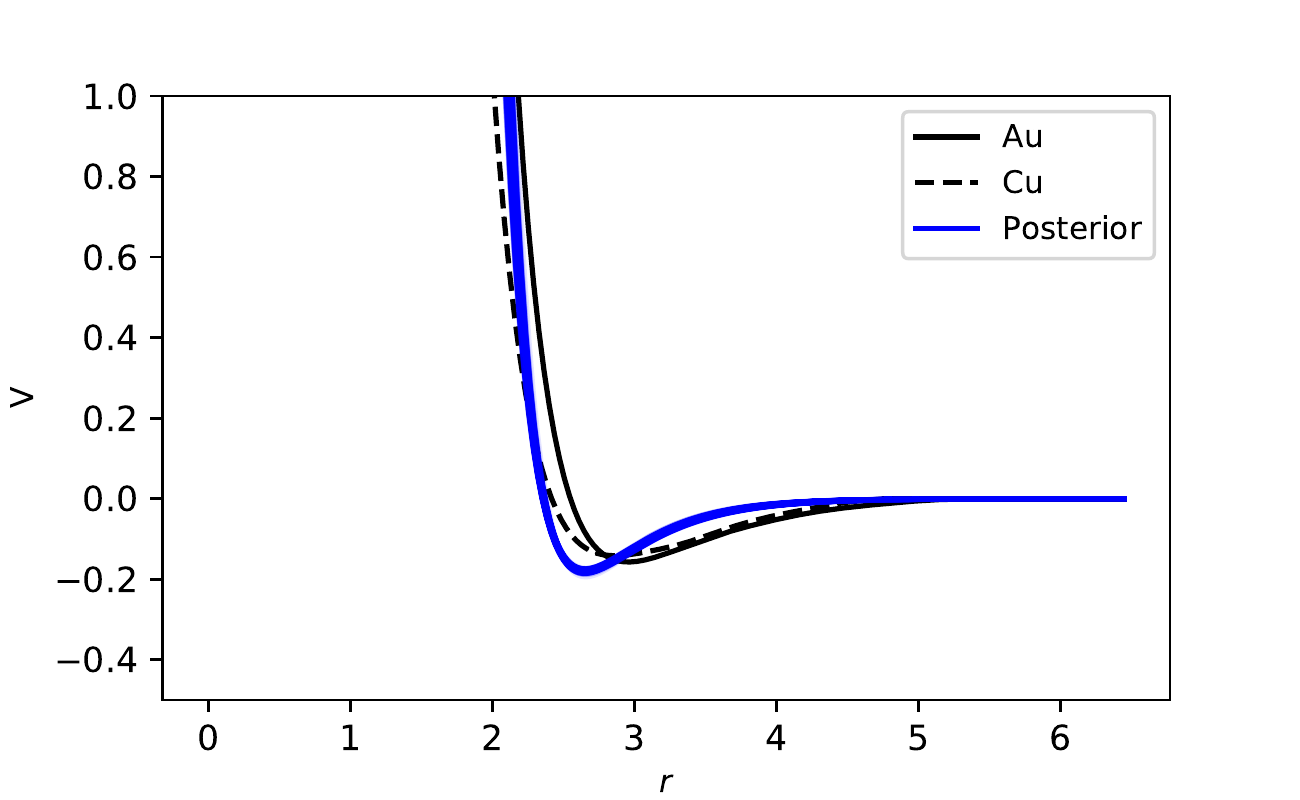}
    \caption{1000 posterior samples (overlaid in blue) of the pair potential $V_{AB}(r)$ compared to the pure elemental pair potentials.}
    \label{fig:V}
\end{figure}

The strategy presented in \Cref{sec:strategy} begins by selecting an initial box in the parameter space. How to select such a region often depends on a combination of domain scientist opinion and optimization. We started with a hypothesis that the potential model should be bounded in some sense by the baseline pure elemental potentials. This logic was readily implementable for the pair potential $V_{AB}(r)$ in \Cref{eq:morse}. Since the pure elemental pair potentials $V_{AA}(r)$ and $V_{BB}(r)$ had graphs of the Morse shape (see \Cref{fig:V}), it was possible to extract effective values for $r_{eq}$, $D$, and $\alpha$ and use them to constrain the corresponding model parameters. We had limited intuition regarding the electron density parameters $S_A$ and $S_B$ and so they were left unconstrained. We identified a best fit parameter vector subject to these constraints. We then examined hyperrectangles of incrementally larger sizes centered about this point. The initial surrogate domain was chosen to be the largest that had acceptable surrogate modeling error for all QoIs. What was deemed ``acceptable'' here depended on the QoI. For example, we allowed for mean relative errors of up to $5-10\%$ for lattice parameters and the elastic constants. Mixing enthalpy tended to have larger relative error as it can take on values near zero. We also note that in our iterative inference process, the surrogates (and the corresponding surrogate domain) were continually refined and improved, leading to the results we discuss below.  

Throughout the various stages of the analysis, the surrogate domain was manually adapted so as to balance both the surrogate fitting accuracy and the confinement of resulting posterior samples. Similarly, the prior was also updated when necessary to ensure that the surrogate domain was contained. We employed uniform priors on the model parameters $\theta$ and uninformative priors on the error parameters $\ln \sigma$ and $\ln \tau$. Samples were drawn via an Adaptive Metropolis MCMC (AM-MCMC) algorithm implemented in UQTk \cite{debusschere2017uqtk,debusschere2004numerical}. In the final run, AM-MCMC was applied for $10^6$ steps and produced an acceptance fraction close to $0.25$. Each parameter chain had an integrated auto-correlation length of roughly $25$ steps, suggesting that the chain has a statistical efficiency equivalent of approximately $4 \times 10^4$ independent samples \cite{hogg2018data,liu2008monte}. Moreover, these results were stable for different chain initializations as well as minor enlargements of the prior domain, which, we recall, was used to ensure the analysis did not stray too far from where the surrogate models were trained. Both statistics suggest that the posterior distribution was sufficiently explored. 

\begin{figure}[hbtp]
    \centering
    \includegraphics[clip,trim=0 0 0 0mm,width=0.75\textwidth]{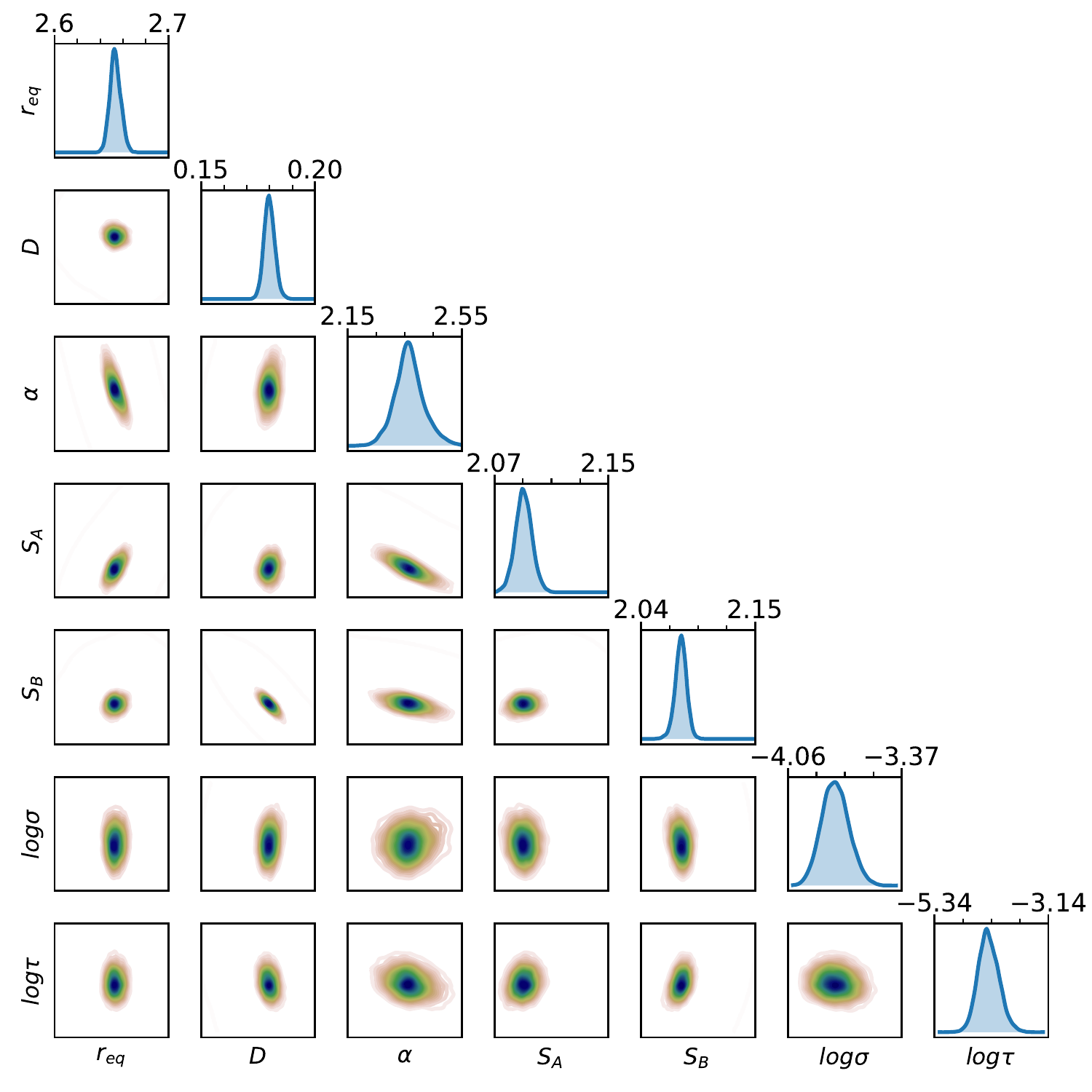}
    \caption{One- and two-dimensional marginal posterior distributions on model parameters and hyperparameters.}
    \label{fig:marginals}
\end{figure}

The resulting one- and two-dimensional (1D/2D) marginal posterior distributions, on both model parameters and model error hyperparameters, are shown in \Cref{fig:marginals}. The axis limits for the model parameters have been set to match the prior domain, which in this case is equivalent to the training region for the surrogates. We find unimodal 1D/2D marginal posteriors, with near-Gaussian 1D marginals. Maximum \emph{a posteriori} (MAP) and mean parameter values, as well as their associated standard deviations, are reported in \Cref{tab:parameter_stats}. We note that the resulting uncertainties are small, as the largest coefficient of variation, being the ratio of the standard deviation to the mean, is about 2\%.

The 2D marginal posteriors reveal several pairwise correlations among parameters. For example, the pairs $(\alpha, S_A)$, $(\alpha, S_B)$, and $(D,S_B)$ all exhibit negative correlations. Recall that these parameters appear in different components of our model -- with $D$ and $\alpha$ belonging to the Morse potential and $S_A$ and $S_B$ belonging to the elemental electron densities. The presence of correlations is a manifestation of the coupling among model parameters in fitting data for the selected QoIs. Accounting for uncertain parameter correlations provides for more accurate estimation of uncertainties on model predictions, avoiding the over-prediction of output uncertainties that would ensue from neglecting such correlations.
\FloatBarrier
\begin{table}[h!]
  \begin{center}
    \caption{MAP/mean parameter values, and their standard deviations}
    \label{tab:parameter_stats}
    \begin{tabular}{c|c|c|c|c|c}
    \hline
       & $r_{\rm eq}$ & $D$ & $\alpha$ & $S_{\rm A}$ & $S_{\rm B}$ \\
      \hline
      MAP (mean) & $2.65$ ($2.65$) & $0.18$ ($0.18$) & $2.36$ ($2.37$) & $2.09$ ($2.09$) & $2.08$ ($2.08$)\\
      $\sigma$ & $0.0048$ & $0.0024$ & $0.051$ & $0.0059$ & $0.0056$\\
    \end{tabular}
  \end{center}
\end{table}
We note that after several stages of the inference procedure, the bulk of the posterior had shifted outside the initial domain. \Cref{fig:V} illustrates this by showing that the final set of posterior samples for the pair potential $V_{AB}(r)$ fall outside the ranges described by the pure elements over a part of the range of $r$. For example, the parameter $r_{eq}$, which specifies the location of the minimum of $V_{AB}(r)$, was initially confined between the two minima associated with the pure elements. The marginal posterior samples for this parameter, however, are all below this range. A similar observation can be made about the parameter $D$, which controls the depth of the minimum. Recall from \Cref{sec:strategy} that the initial surrogate domain defined above was essentially set to facilitate construction of accurate surrogates. Similarly, the prior range, which subsumes the surrogate domain, was designed to focus the sampling in the vicinity of where the surrogates are valid, rather than being based on known prior information or any physical constraints. Hence, we allowed the procedure to freely move outside this initial region in pursuit of a better model fit. 
\FloatBarrier
In \Cref{fig:surrogate_latticePar,fig:surrogate_mixEnthalpy,fig:surrogate_C11,fig:surrogate_C12,fig:surrogate_C44,fig:surrogate_bulkMod}, we show corresponding predictive distributions using the surrogate models. In particular, and recalling the formulation in \Cref{sec:formulation}, we highlight the pushforward posterior (PFP), \Cref{eq:pfp}, which is the density on the physical/forward model output $\textbf{f}(\theta)$ given the posterior density on its parameters $\theta$, and the posterior predictive (PPD), \Cref{eq:ppd}, which is the density on the data model output $\textbf{f}(\theta) + \epsilon$ given the posterior density on its parameters $\lambda=(\theta,\ln\sigma,\ln\tau)$. The MAP predictions are included as a visual single-point reference. 
Similarly, we also incorporate the deviation from the linear rule of mixtures for both the DFT data and the MAP estimate in the inset plots for all of the QoIs except mixing enthalpy (for which the linear rule of mixtures leads to the zero function). Note that this rule provides a simple empirical relationship for the composition dependence of a physical property based on the pure elements and hence provides a reference point for the predictions. 
Additionally, note that these distributions were generated by subsampling $1000$ parameter values from the MCMC chain and evaluating the corresponding $102$ QoIs with the surrogates. 
\FloatBarrier
\vspace{-15pt}
\FloatBarrier
\vspace{-15pt}
\FloatBarrier
\vspace{-15pt}
\FloatBarrier
\vspace{-15pt}
\FloatBarrier
\vspace{-15pt}
\FloatBarrier
\Cref{fig:surrogate_latticePar}-\Cref{fig:surrogate_bulkMod} highlight how the indicated physical properties vary with composition. Recall, the properties for the pure elements ($0\%$ and $100\%$ Au) are essentially fixed inputs to the model. The normalization of the DFT data, described in \Cref{sec:rampage}, is performed to ensure that the LAMMPS and DFT QoIs exactly match at these end points. 

There are various observations to be made about the uncertain predictions in these figures. To begin with, we note that the MAP predictions agreement with DFT data varies depending on the model output and mixture conditions. Clearly agreement is generally better for the lattice parameter, $C_{12}$, $C_{44}$, and bulk modulus, with worse agreement for mixing enthalpy and $C_{11}$. This is not unexpected given the simplicity of the EAM model and the range of complexity in the QoIs. Further, the results show that the PFPs are quite narrow, especially in the case of lattice parameters. This feature is consistent with the small posterior parametric uncertainties presented above, whereas the PPDs are much wider. This is expected, of course, since the PPD also incorporates the model error. One important observation, from a diagnostic perspective, is that the PPD widths generally span the scatter of the DFT data, which indicates that the formulation underlying the inference (i.e., \Cref{eq:probmodel}) is reasonable in that the data appear as plausible outputs from the calibrated data model. We note, however, that there are a few outliers -- namely, $C_{11}$ at $50\%$ gold and $C_{44}$ at $56\%$ gold. Nonetheless, the key finding from a predictive perspective is the degree of disagreement between the calibrated IAP model predictions, as exhibited by the PFPs, and DFT data.

Aside from the obvious path forward of developing a better IAP model, which is outside the scope of the present work, it is useful to consider what quantified uncertainties can add to qualify predictions from our existing imperfect model. Clearly, the PFP uncertainty is small relative to the disagreement with the DFT data and thus the associated confidence in predictions is overly optimistic. Moreover, employing more data for fitting would only reduce this uncertainty further, thereby enhancing confidence in predictions from an imperfect model~\cite{sargsyan2015statistical,sargsyan2019embedded}. From a modeler's perspective, this is clearly undesirable. Rather, given the unavoidable imperfections in a physical model, we submit that it would be useful for the associated \emph{uncertain} predictions to provide a reliable measure of the disagreement from the truth (i.e., the DFT calibration data in the present context).

In Bayesian calibration, this can be accomplished by augmenting the physical model with a model error term, i.e., a statistical model that captures the discrepancy between the best fit physical model predictions and the data~\cite{kennedy2001bayesian,sargsyan2015statistical,sargsyan2019embedded}. In our present setting, this is provided by viewing the PPD as the predictive model. As evidenced by the results above, the PPD uncertainties are far more reliable as conservative measures of how far-off the model predictions can be from the truth. In this way, UQ provides means for making predictions under realistic/imperfect conditions with uncertainty that captures the actual degree of discrepancy. It is worth noting, of course, that the structure of the specific model error is itself a modeling choice, and can include \emph{e.g.} additive Gaussian processes~\cite{kennedy2001bayesian}, or embedded~\cite{sargsyan2015statistical,sargsyan2019embedded} formulations. 

An important lead-up to the following section is that model error constructions built on top of the physical model output as we have effectively done here, and as in~\cite{kennedy2001bayesian}, are only useful for informing calibration QoIs, and do not claim to capture model error related uncertainty in other model outputs not present in the calibration. We emphasize this particular aspect in the next section where we use our calibrated IAP model to predict interatomic forces, a QoI not included during calibration. As will be shown, these predictions are only of the PFP type as our output model error does not translate to the new setting. On the other hand, we note that embedded constructions for model error ~\cite{sargsyan2015statistical,sargsyan2019embedded, Huan:2018b} can be used to enhance uncertainty in other model predictions, albeit without any effective control (as we have in calibration) of the degree to which those predictive uncertainties capture the discrepancy from the truth. 

Having calibrated the IAP model using surrogates, one important question is how reflective are the above results of the actual LAMMPS simulation? As evidenced by the figures in \Cref{sec:appendix_surrogate_error}, which show the surrogate modeling error, the surrogates provide very close approximations to the LAMMPS code over the identified posterior region. Moreover, the LAMMPS-based predictive distributions appear virtually identical to the surrogate predictions shown in \Cref{fig:surrogate_latticePar,fig:surrogate_mixEnthalpy,fig:surrogate_C11,fig:surrogate_C12,fig:surrogate_C44,fig:surrogate_bulkMod}. Hence, the validity of our above conclusions follows for the LAMMPS simulation. 

\subsection{Predicting forces with the calibrated model}
\label{sec:caseStudy_forces}

Models are often used to make predictions in settings different from those during calibration. This is justified if the physical principles behind the model are applicable in both the calibration and prediction domains. In this section, we examine the performance of our calibrated IAP model in predicting interatomic forces -- a QoI not explicitly included during calibration. We note that predictions in this setting are all of the PFP type, i.e. as in \Cref{eq:pfp2}. As stated previously, uncertain predictions incorporating our above model error construction and the associated PPDs, are only relevant to the calibration QoIs. This presents a challenge as \Cref{sec:caseStudy_calibration} established that the PFPs are overly confident on the calibration data and do not convey the appropriate degree of uncertainty. The results below similarly reflect this. Moreover, these results highlight the need for a more flexible IAP model and echo the preceding discussion on embedded forms of model error. 

Before the advent of force matching to DFT data for IAP fitting~\cite{ercolessi1994interatomic}, fitting data consisted almost exclusively of lattice parameters, cohesive energies, and elastic constants~\cite{voter1994embedded}. Although (cohesive) energies and forces are direct outputs of the IAP, we did not incorporated them in the calibration detailed in \Cref{sec:caseStudy_calibration}. Moreover, the elemental IAPs for Au and Cu that are input to the model were not force-matched in their construction \cite{zhou2004misfit}. Calibration with force data presents significant associated complexity beyond energy-fitting, even in a deterministic framing, and is best reserved for future work in the Bayesian setting. Recall that our $5$-parameter IAP was calibrated with a collection of physical properties from $17$ different compositions, all at a temperature of $0$~K. For a given composition, these physical properties are characteristics of the associated minimum energy configuration. Specifically, the lattice parameter and mixing enthalpy are determined by the arrangement of atoms and their energy at the ground state. The elastic constants elicit second order information about this ground state, i.e., are related to the curvature of the energy surface. In contrast, forces on particular atoms are computed by taking the negative gradient of the potential energy surface. Indeed at the minimum energy configuration, these forces are zero. Once atoms are displaced from their ground state positions, e.g., by thermal motion, the forces are no longer negligible. The specific question we ask in this section is the following: having calibrated our model on the mentioned physical properties following the traditional paradigm for fitting~\cite{voter1994embedded}, how predictive is it of interatomic \textit{forces} for systems at temperatures of $300$~K, $750$~K, and $1000$~K?

\begin{figure}[hbtp]
\centering
\input{force_workflow.tex}
\caption{Molecular dynamics simulation workflow for computing forces.}
\label{fig:force_workflow}
\end{figure}
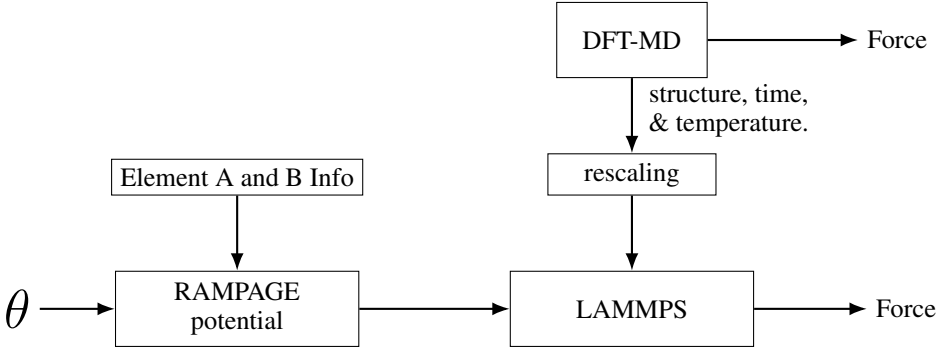

The workflow for computing forces is illustrated below in \Cref{fig:force_workflow}.
\FloatBarrier
This setup is driven by a collection of DFT-based molecular dynamics simulations. Note that the DFT generated structures are rescaled based on a linear rule of mixtures, similar to \Cref{eq:ruleofmixtures} for lattice constants, before being input into LAMMPS. More details can be found in \Cref{sec:appendix_dft-md}.

Note that how we visualize the results here differs from the calibration section due to the amount of force data output by the simulation. In the previous section, the calibration data could be conveniently organized to illustrate the pushforward posteriors for each physical property and composition. Here, for a given parameter vector, composition, and temperature, we now have as output a vector of forces whose dimension extends across a collection of simulated time instances, atom positions, and force components. In what follows, we first present parity plots comparing the IAP/LAMMPS-based and DFT-based force vectors for individual parameter evaluations -- specifically the MAP estimate. We then present the pushforward posteriors for certain error metrics computed from the corresponding parity plots. These results emphasize the disparity in force predictions between our calibrated model's PFPs and the reference DFT data. 

\begin{figure}[hbtp]
     \centering
     \begin{subfigure}[b]{0.49\textwidth}
         \centering
         \includegraphics[trim=0 0 0 0mm, clip]{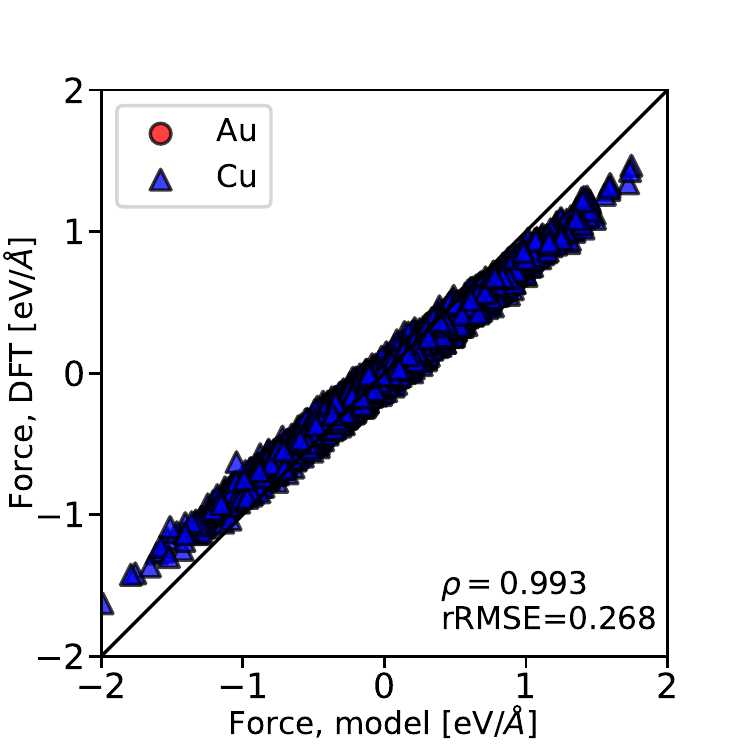}
         \caption{Au (0\%), Cu (100\%) at $\textrm{T} = 300$~K}
         \label{fig:A00B32}
     \end{subfigure}
     \hfill
     \begin{subfigure}[b]{0.49\textwidth}
         \centering
         \includegraphics[trim=0 0 0 0mm, clip]{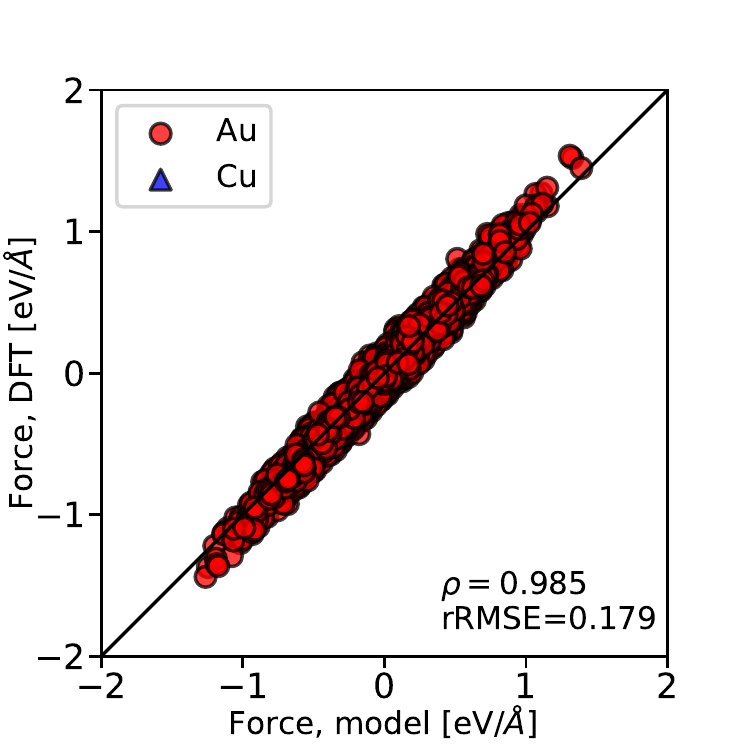}
         \caption{Au (100\%), Cu (0\%) at $\textrm{T} = 300$~K}
         \label{fig:A32B00}
     \end{subfigure}
     \caption{Scatter plots of the force component predictions for the pure elements at a temperature of $\textrm{T} = 300$~K. The points correspond to force components ($x$, $y$, and $z$) and are aggregated across all time steps of the simulation path. The diagonal line corresponds to perfect force matching. The scatter is characterized by the corresponding Pearson correlation coefficient ($\rho$) and the relative root mean square error (rRMSE).}
    \label{fig:pure_parity}
\end{figure}

As with the calibration setup, the pure elemental components of the IAP model are used as a reference in rescaling the DFT structures to ensure the force computations are compatible, as described in \Cref{sec:appendix_dft-md}. Moreover, the elemental potentials enter the model as already fitted and are hence unaffected by the model parameters. As such, the force predictions for these pure elements are expected to be accurate by design and therefore may be used as a benchmark against which to compare the intermediary compositions. The pure elemental parity plots are shown below in \Cref{fig:pure_parity} for a temperature of $\textrm{T} = 300$~K.
\FloatBarrier
The scatter of the data is effectively aggregated over both time and components. We use the Pearson correlation coefficient ($\rho$) and the relative root mean square error (rRMSE) to characterize the prediction error in these results, as indicated in the figure. These metrics are defined as follows:
\begin{equation}
    \label{eq:force_metrics}
    \begin{aligned}
    & \rho = \frac{ 
    \underset{j \in \mathcal{J}}{\sum} 
    \left(F_{j}^{\ \textrm{model}} - \overline{F}^{\ \textrm{model}} \right) \left(F_{j}^{\ \textrm{DFT}} - \overline{F}^{\ \textrm{DFT}}\right)
    }{\sqrt{\underset{j \in \mathcal{J}}{\sum} 
    \left(F_{j}^{\ \textrm{model}} - \overline{F}^{\ \textrm{model}} \right)^2 }\sqrt{\underset{ j \in \mathcal{J}}{\sum} 
    \left(F_{j}^{\ \textrm{DFT}} - \overline{F}^{\ \textrm{DFT}} \right)^2}} \\
    & \textrm{RMSE} = \sqrt{ \frac{1}{|\mathcal{J}|}\underset{ j \in \mathcal{J}}{\sum} 
    \left( F_{j}^{\ \textrm{model}} - F_{j}^{\ \textrm{DFT}}\right)^2} \\
    &\textrm{rRMSE} = \frac{\textrm{RMSE}}{\sqrt{ \frac{1}{|\mathcal{J}|}\underset{ j \in \mathcal{J}}{\sum} 
    \left( F_{j}^{\ \textrm{DFT}}\right)^2}},
    \end{aligned}
\end{equation}
where all summations are taken over time steps, atoms, and components, and
\begin{equation}
\begin{aligned}
& \overline{F} = \frac{1}{|\mathcal{J}|}\underset{ j \in \mathcal{J}}{\sum} F_{j} \\ 
& \mathcal{J} = \{ (t,a,c) : t\in \textrm{time}, \ a\in \textrm{atoms}, \ c\in \{x,y,z\} \} .
\end{aligned}
\end{equation}
In the ideal case, the data would fall exactly on the diagonal, i.e., a line of slope $1$ and intercept $0$ ($\rho = 1$, $\textrm{RMSE}=0$). The above plots show that the force predictions for both pure Au and pure Cu are reasonably accurate. 

\begin{figure}[hbtp]
     \begin{subfigure}[b]{0.4\textwidth}
         \centering
         \includegraphics{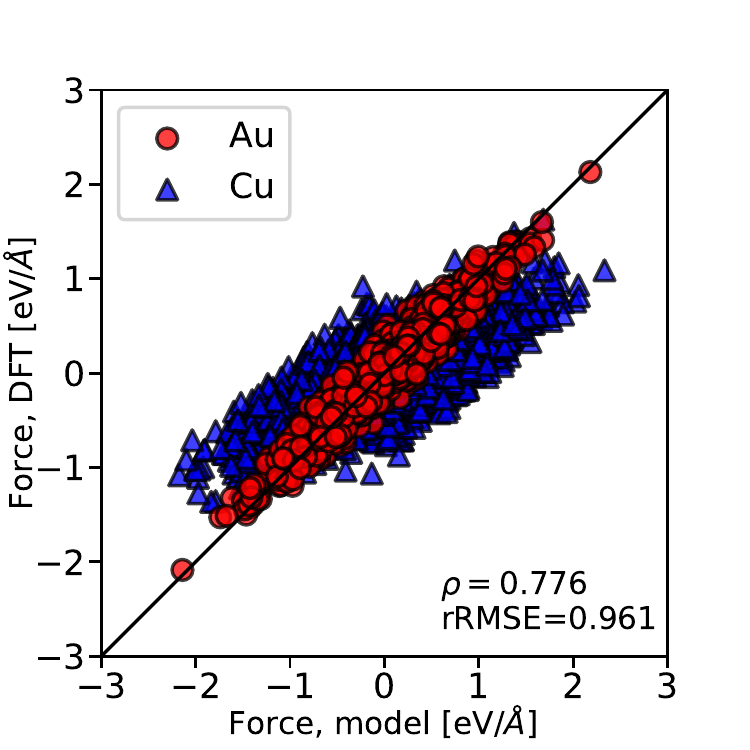}
         \caption{Au (25\%), Cu (75\%) at $\textrm{T} = 300$~K}
         \label{fig:A027B081}
     \end{subfigure}
     \hfill
     \begin{subfigure}[b]{0.4\textwidth}
         \centering
         \includegraphics{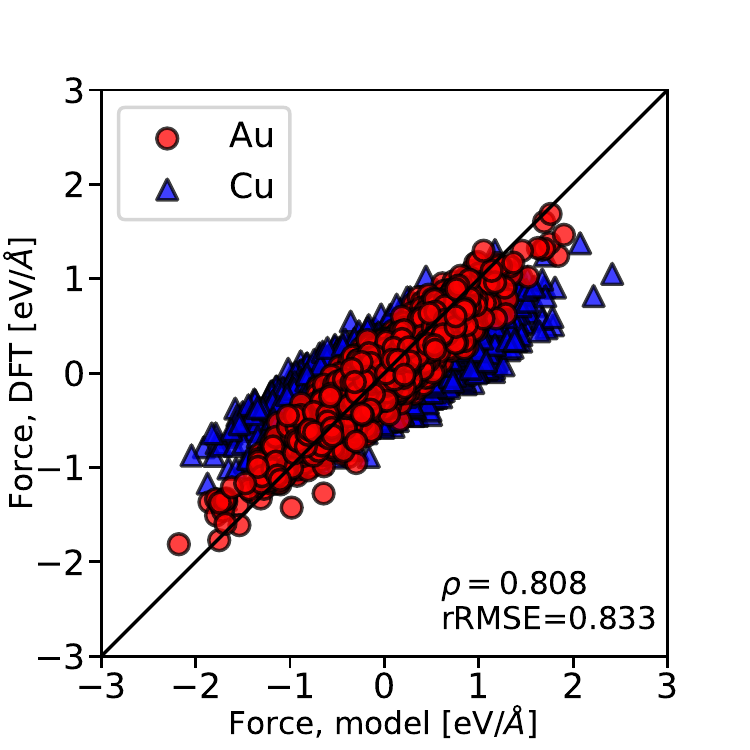}
         \caption{Au (50\%), Cu (50\%) at $\textrm{T} = 300$~K}
         \label{fig:A054B054}
     \end{subfigure}
     \vspace{1em}
     \begin{subfigure}[b]{\textwidth}
         \centering
         \includegraphics[width=0.4\textwidth]{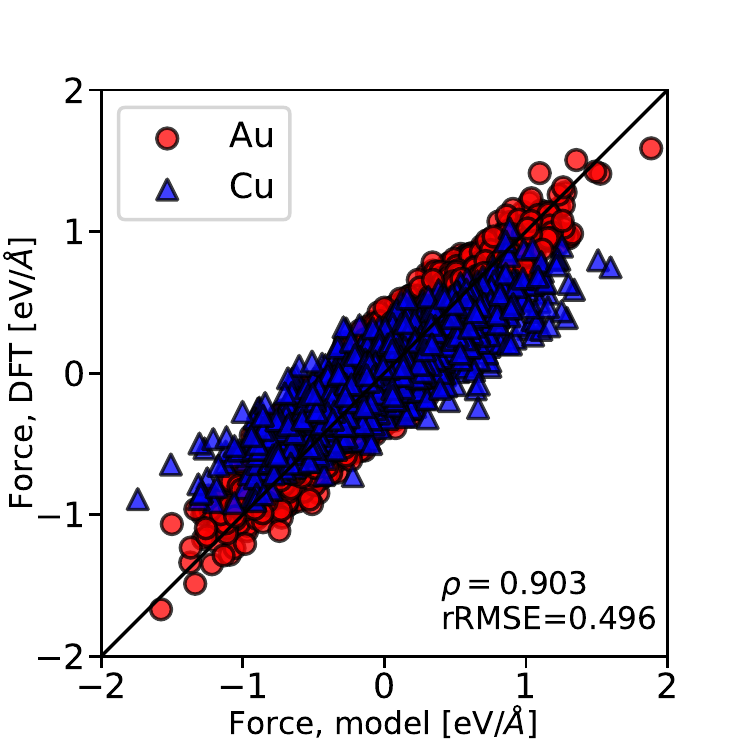}
         \caption{Au (75\%), Cu (25\%) at $\textrm{T} = 300$~K}
         \label{fig:A081B027}
     \end{subfigure}
    \caption{Scatter plots of force component predictions for the MAP parameter estimate at $\textrm{T} = 300$~K. The points correspond to force components ($x$, $y$, and $z$) and are aggregated across all time steps of the simulation path. The diagonal line corresponds to perfect force matching.}
    \label{fig:map_parity}
\end{figure}

Having established the pure elements as benchmarks, we turn to the actual IAP model performance on the intermediary compositions. Parity plots associated with just the MAP parameter estimate are displayed below in \Cref{fig:map_parity} for compositions $25\%$, $50\%$, and $75\%$ Au at a temperature of $T=300$~K.
\FloatBarrier
The error metrics show that force predictions degrade as compared to the pure elements. Intriguingly, the element types appear quite distinctly and several characteristics of the pure elemental potentials are retained. Au atoms tend to be scattered along the parity line, whereas Cu atoms are scattered around a line with slope less than one. This perhaps reflects how the RAMPAGE model fuses and reuses the two elemental potentials. The relatively wide scatter also leads to situations where the LAMMPS predicts forces with opposite sign to the DFT values, as evidenced by samples falling in the second and fourth quadrants of the figures.

\begin{figure}[hbtp]
     \centering
     \includegraphics{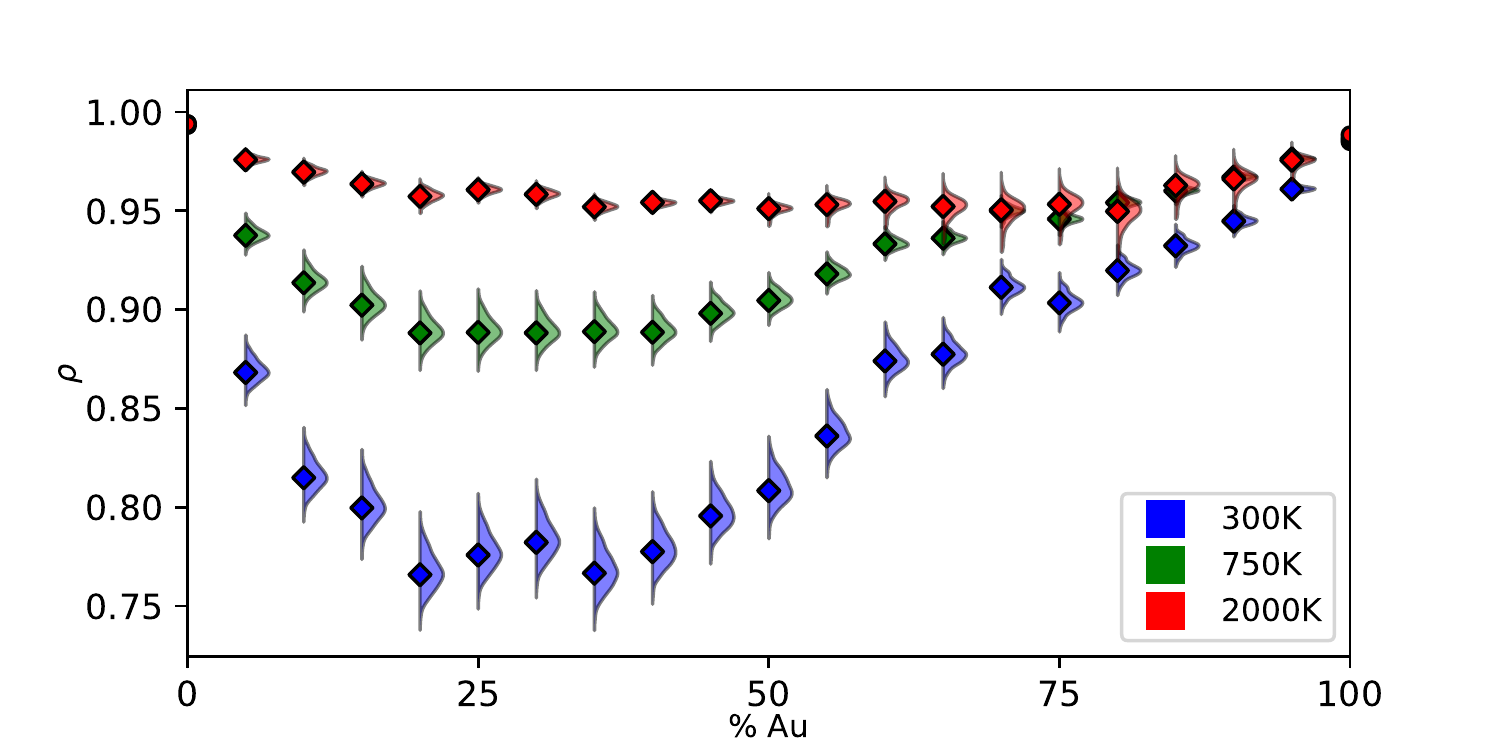}
     \caption{Pushforward posterior distributions for the correlation coefficients. The diamond markers indicate the MAP estimate.}
     \label{fig:PFP_cor}
 \end{figure}

\begin{figure}[hbtp]
 \centering
 \includegraphics{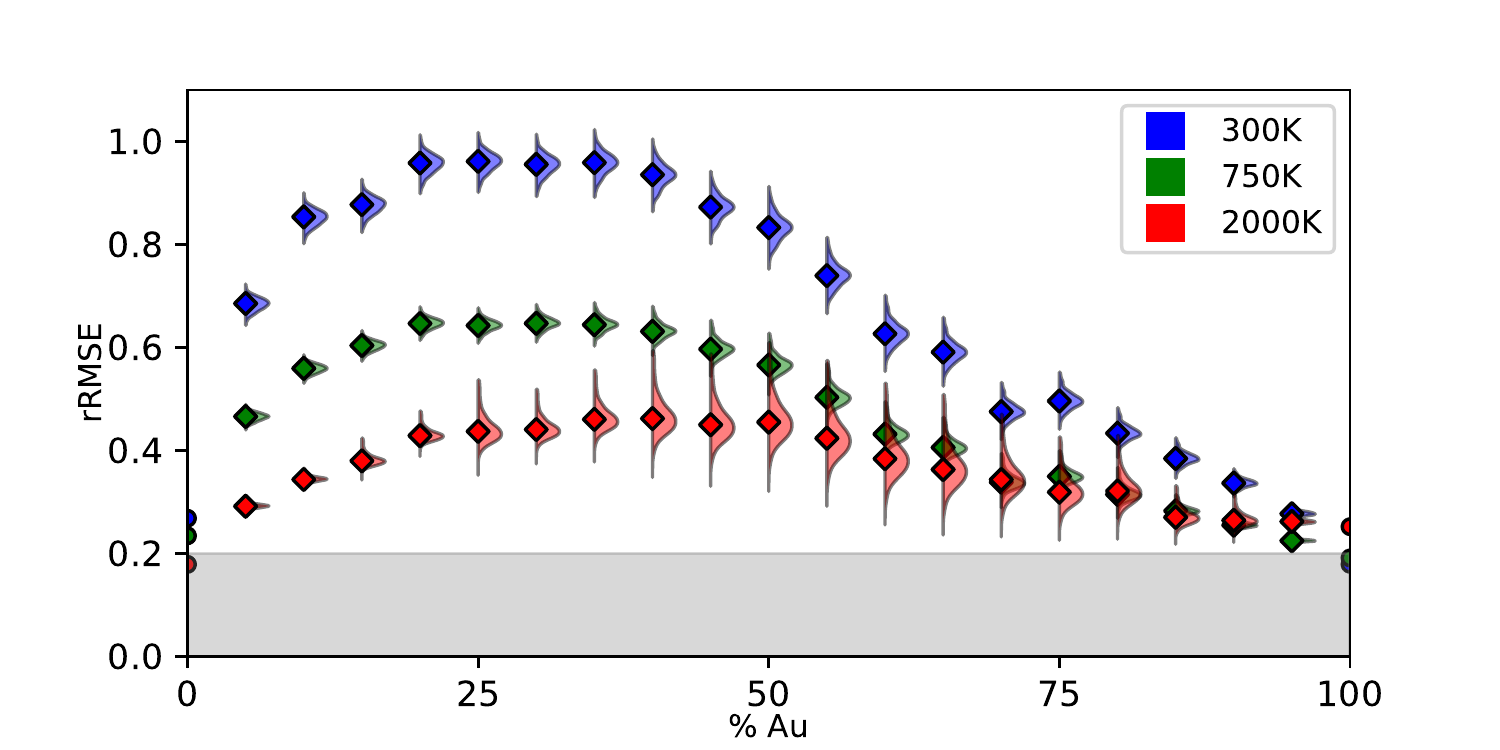}
 \caption{Pushforward posterior distributions for the relative root mean square errors (rRMSE). The diamond markers indicate the MAP estimate. For reference, the shaded region below rRMSE values of 0.2 contains the calibration QoI errors. }
 \label{fig:PFP_rRMSE}
\end{figure}

Since the error metrics provide useful summaries of the force predictions for given parameter vectors, we show their corresponding PFPs in \Cref{fig:PFP_cor} and \Cref{fig:PFP_rRMSE}. Note that in the latter figure, we have also superimposed the relative error range of the calibration QoIs for comparison. More details on these calculations can be found in \Cref{sec:appendix_calibration_rrmse}.
\FloatBarrier
The PFPs show the model performance over the marginal posterior distribution $p(\theta|\bf{y})$, rather than just the single MAP parameter illustrated previously. Evidently, the force predictions tend to be best at higher temperatures, although they do not reach the quality of the pure elements or the calibration QoIs. The narrowness of these PFPs reflects our earlier discussion for the calibration QoIs regarding the differences between the PFPs and PPDs  -- the posterior uncertainty in our model parameters do not convey the appropriate level of uncertainty. For reference, we note that in the ideal case of a perfect model, we would expect to find narrow predictive distributions that contain $\rho=1.0$ and ${\rm rRMSE} = 0.0$. In the present context, and without the benefit of model error, the PFPs for the force field data show a relatively larger departure from the corresponding DFT force field data compared to results shown in the previous section. This fact is illustrated by the shaded gray region in \Cref{fig:PFP_rRMSE} and further documented in \Cref{sec:appendix_calibration_rrmse}.

The results display a number of interesting trends. For example, the force predictions tend to be worst at the lower end of the composition spectrum, in which the system contains a greater proportion of copper atoms. Similarly, the differences among temperatures are also greater for systems dominated by copper atoms. The transitions in \Cref{fig:PFP_cor} and \Cref{fig:PFP_rRMSE} between the pure copper system ($0\%$ Au) and the first mixture ($5.6\%$ Au) are far steeper than that between the last mixture ($94.4\%$ Au) and pure gold. In comparison, such composition-dependent asymmetries are not nearly as apparent in the calibration QoIs -- mixing enthalpy perhaps being the only example. In general, higher temperatures are associated with correlation coefficients closer to 1, i.e., the model and DFT predictions become highly correlated. 

Beyond what was stated previously with respect to model error, these results suggest that including forces and energies as calibration QoIs, e.g., as done in the force matching method \cite{ercolessi1994interatomic}, would be a useful next step in developing a model with improved predictive performance. 

\section{Conclusions}

Recent growth in the availability of computing resources has made uncertainty quantification in atomistic simulation a more promising endeavor. In the current work, we have presented a Bayesian calibration strategy for fitting RAMPAGE alloy potentials for an Au-Cu system. The RAMPAGE methodology constructs potentials for multi-component systems by fusing together existing single-element potentials in a coherent manner. The modeling effort is then solely expended on characterizing interactions among the different element types. In this setting, calibration requires fitting to DFT data from a series of compositions, spanning the range from a pure Cu system to a pure Au system. Direct application of Bayesian inference techniques, however, lead to computational challenges. To mitigate these difficulties, we employed a sequential approach based on local inference steps in order to analyze a $5$-parameter RAMPAGE potential model. Using this example, we demonstrated how uncertainty quantification and its associated predictive distributions can provide more complete assessments of a model's quality and its limitations. In particular, we highlighted the important role played by model error when working with an imperfect model. Our results motivate exploring more recent and powerful forms of model error and suggest avenues for model improvement, such as the inclusion of force and energy data during calibration.     

\section*{Acknowledgement}
This work was supported in part by the U.S. Department of Energy, Office of Science, Office of Advanced Scientific Computing Research, Scientific Discovery through Advanced Computing (SciDAC) program through the FASTMath Institute. This research used resources of the National Energy Research Scientific Computing Center (NERSC), a U.S. Department of Energy Office of Science User Facility located at Lawrence Berkeley National Laboratory, operated under Contract No. DE-AC02-05CH11231. Further resources were used at The Ohio Supercomputer Center, Project No.\ PAS0072. EW and WW acknowledge funding from the Air Force Office of Scientific Research (AFOSR) under Contract No.~FA9550-19-1-0378. Sandia National Laboratories is a multimission laboratory managed and operated by National Technology \& Engineering Solutions of Sandia, LLC, a wholly owned subsidiary of Honeywell International Inc., for the U.S. Department of Energy's National Nuclear Security Administration under contract DE-NA0003525. This paper describes objective technical results and analysis. Any subjective views or opinions that might be expressed in the paper do not necessarily represent the views of the U.S. Department of Energy or the United States Government.
\appendix
\section{Appendix}
\subsection{\textit{Ab initio} Computational Methodology}
\label{sec:appendix_comp}

\subsubsection{Density Functional Theory Calculations}
\label{sec:appendix_dft}

All DFT calculations were performed within the Vienna \textit{Ab initio} Simulation Package (VASP) using Projector Augmented Wave (PAW) potentials with Perdew, Burke, and Ernzerhof (PBE) exchange-correlation functional \cite{kresse1994ab,kresse1996efficiency,kresse1996efficient,kresse1999ultrasoft,perdew1997generalized}. Standard DFT calculations were performed with a Monkhorst-Pack k-point mesh with a grid spacing of 0.1~\r{A}$^{-1}$ \cite{monkhorst1976special}. A cutoff energy of 550 eV was chosen for the plane-wave basis. 

Relaxations were performed with Methfessel/Paxton 1st order smearing, followed by a static energy calculation using the tetrahedron method with Blöchl corrections to compute ground state energies. 

To compute the bulk modulus, the Murnaghan equation of state was fitted to 5 DFT calculations performed under isotropic strains ranging from 0.995 to 1.005. The elastic tensor of a cubic crystal contains 9 non-zero components, $C_{11}$, $C_{22}$, $C_{33}$, $C_{12}$, $C_{13}$, $C_{23}$, $C_{44}$, $C_{55}$, and $C_{66}$. Cubic symmetry reduces these 9 non-zero components to 3 unique elastic constants since $C_{11} = C_{22} = C_{33}$, $C_{12} = C_{13} = C_{23}$, and $C_{44} = C_{55} = C_{66}$. Each of the 9 non-zero components of the elastic tensor for cubic symmetry crystals were computed independently using 3 strain states ranging from 0.9975 to 1.0025 by fitting their stress-strain results with Hooke's Law. $C_{11}$, $C_{12}$, and  $C_{44}$ were then obtained by averaging across their identical components to minimize errors resulting from the use of the SQS in the DFT simulations. VASPKIT was used to facilitate and set up calculations. \cite{VASPKIT}

\subsubsection{\textit{Ab Initio} Molecular Dynamics}
\label{sec:appendix_dft-md}

In order to create physically meaningful force sets, it becomes necessary to generate appropriately perturbed structures. This was accomplished by thermalizing the structures with \textit{ab initio} Molecular Dynamics using VASP within the PAW-PBE framework. Unlike classical MD, which relies on fitted IAPs, \textit{ab initio} MD uses electronic structure calculations to compute energies and forces. \textit{Ab initio} MD suffers from poor scaling, thus simulation sizes, and therefore utility as remain limited when compared to classical MD. To generate structures and force sets, an $NVT$ ensemble was used with a Langevin thermostat and friction coefficients of 1 ps$^{-1}$. A 0.3 \r{A}$^{-1}$ Monkhorst-Pack k-point grid was used in conjunction with a 425 eV cutoff energy and an energy stopping criterion of $10^{-5}$ eV. For each elemental and 19 intermediary alloy cells, three separate \textit{ab initio} MD simulations at temperatures of $300$~K, $750$~K, and $2000$~K were performed to generate varying force magnitudes. Each simulation was run for 2000 time steps with $dt=5$ fs.  

The structures generated by $NVT$ \textit{ab initio} MD used volumes determined by the DFT ground state. As discussed in \Cref{sec:rampage}, the DFT computed lattice constants, and therefore the volume of a simulation cell may deviate from the selected elemental IAPs. The volume of the structures must be normalized before they can be used within the classical MD framework. Otherwise they will 'appear' strained to the elemental IAPs, and result in a systematic deviation in forces that is solely a result of the volume mismatch. Thus, to avoid probing separate locations of the DFT and IAP energy landscapes, the lattice constants of the thermalized structures are normalized according to the rule of mixtures shown in \Cref{eq:ruleofmixtures}. 

\subsubsection{Special Quasirandom Structure generation} 
\label{sec:appendix_sqs}

Solid solution alloys were approximated within DFT, \textit{ab intito} MD, and empirical IAP based MD using face centered cubic special quasirandom structures (SQSs) generated with the ATAT code package \cite{vandewalle2013efficient, von2010generation}. These structures are created using a Monte Carlo algorithm to most closely reproduce the cluster correlation functions of perfectly random substitutional alloys within the periodic boundary conditions imposed in these three atomistic simulation frameworks. All SQSs were generated with pair and triplet correlation cutoffs that included the third nearest neighbor shells. SQSs used for DFT were limited to 32 atom 2$\times$2$\times$2 FCC super cells due to the computational expense associated with the method. 108 atom 3$\times$3$\times$3 FCC super cells were utilized in \textit{ab initio} and empirical MD simulations to validate forces. All calculations of QoIs within LAMMPS utilized 12$\times$12$\times$12 FCC 6912 atom cells. 

\subsection{Surrogate modeling error on the posterior samples}
\label{sec:appendix_surrogate_error}
\begin{figure}[hbtp]
     \centering
     \includegraphics{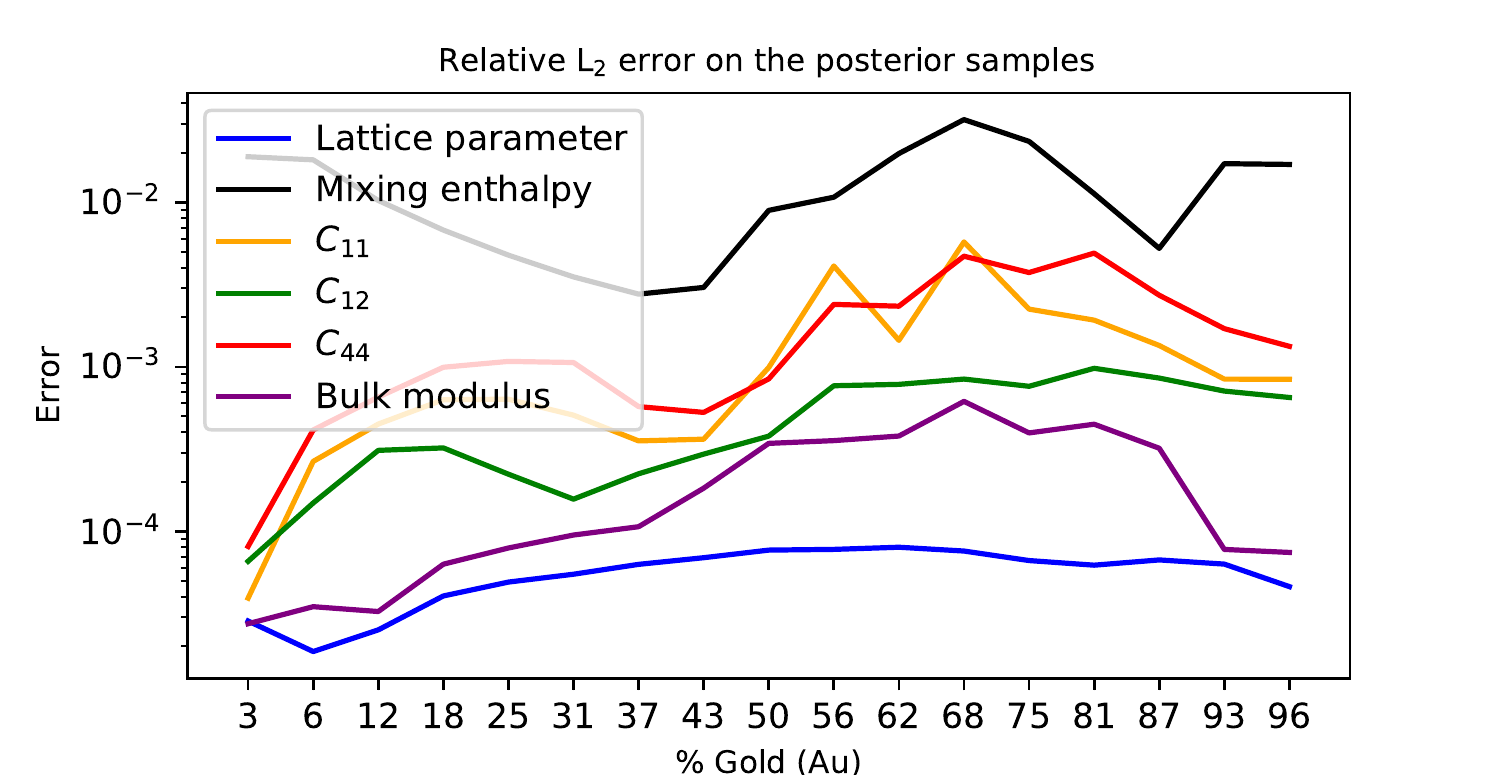}
     \caption{Relative ${\rm L}_2$ error for the indicated QoIs}
     \label{fig:surrogate_error}
\end{figure}

As discussed in \Cref{sec:caseStudy_calibration}, the fidelity of our conclusions depends on the accuracy of the $102$ surrogate models. Below, we assess the error of the surrogate models on the $1000$ posterior parameter samples used in the main text. These samples were not included in the training of the surrogates. For the six physical properties, we report the \textit{relative $\rm L_2$ error} which is computed by, 
\begin{equation}
    \label{eq:MAPE}
   \frac{\sqrt{\sum_{i=1}^{m}  \left(y_{\rm SURR} (\theta_i) - y_{\rm LAMMPS} (\theta_i)\right)^2}}{\sqrt{\sum_{i=1}^{m}  y_{\rm LAMMPS} (\theta_i)^2}},
\end{equation}
where $y_{\rm SURR}$ refers to the surrogate model for a particular QoI, $y_{\rm LAMMPS}$ refers to the corresponding LAMMPS evaluation, and $\{\theta_i\}_{i=1}^m$ are the posterior samples. \Cref{fig:surrogate_error} displays these errors.
\FloatBarrier
In all cases, the relative error is less than $3\%$. Moreover, as noted in the main text, the predictive distributions generated using LAMMPS are indistinguishable from the surrogate-based results. 

\subsection{rRMSEs for the calibration QOIs}
\label{sec:appendix_calibration_rrmse}

In the calibration (\Cref{sec:caseStudy_calibration}) and force testing (\Cref{sec:caseStudy_forces}), different methods were used for visualizing the prediction errors due to differences in the simulation setup. For calibration, we showed the predictive distributions of the physical properties (lattice parameter, mixing enthalpy, etc.) for each composition and physical property. For force prediction, we instead showed parity plots as well as pushforward posteriors of certain error metrics (correlation coefficients and relative root mean square errors (rRMSEs)) for each composition and temperature. A reasonable question to ask is: how would the calibration QoIs look if they were presented in the same manner as the forces?

\begin{figure}[hbtp]
     \centering
     \includegraphics{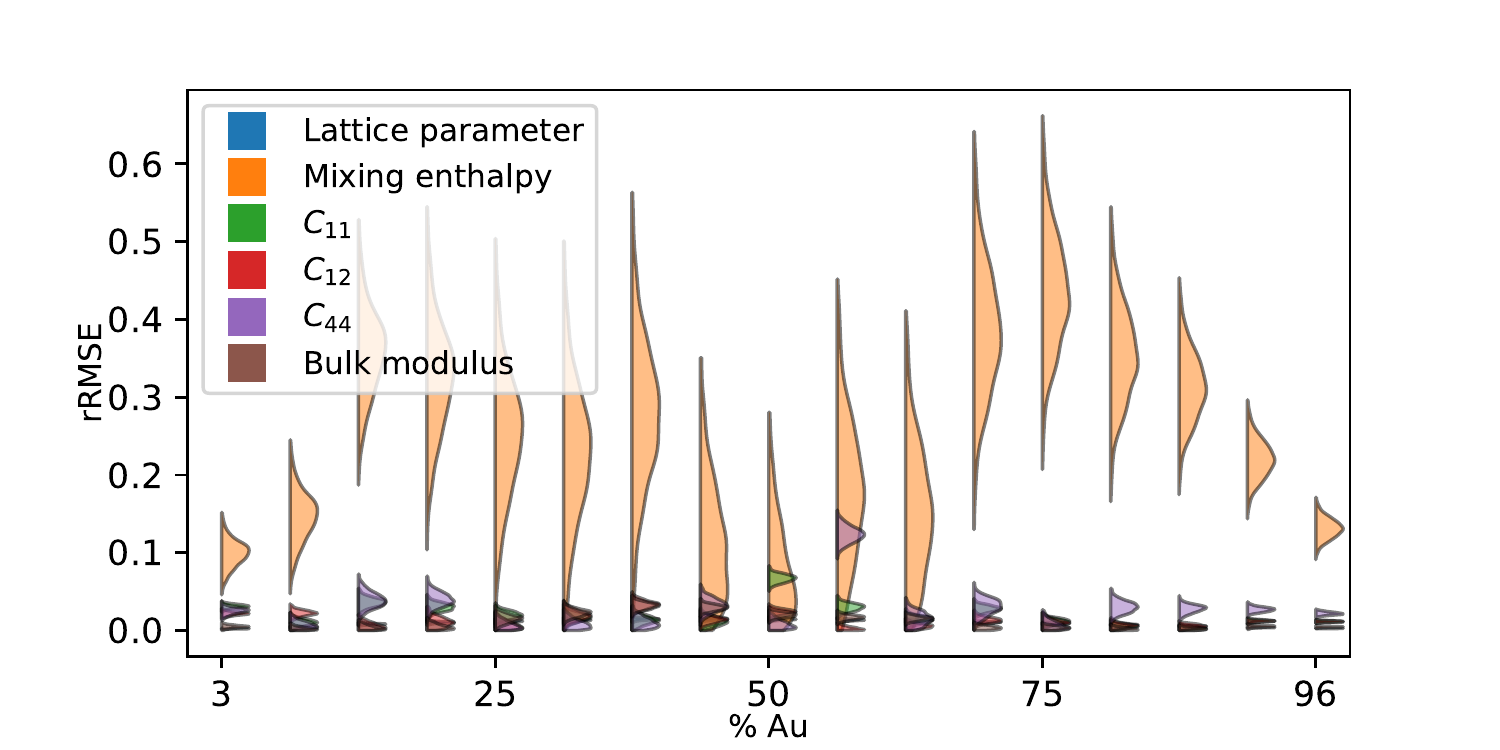}
     \label{fig:rRMSE_calibration}
     \caption{Pushforward posterior distributions for the relative prediction error on the calibration QoIs}
\end{figure}

Since for each composition and physical property the LAMMPS outputs and DFT data are scalars, the corresponding correlation coefficients (and parity plots) are uninteresting. In this case, the rRMSE reduces to just point-wise relative errors, i.e.
\begin{equation}
    \label{eq:rel_error}
    \left| \frac{f_i(x;\theta_k) - y_i(x)}{y_i(x)} \right|,
\end{equation}
where $\theta_k$ is a parameter vector and we have used the same notation as in \Cref{eq:probmodel}. In \Cref{fig:rRMSE_calibration}, we show the pushforward posterior for the relative errors for each physical property and composition. Note that since the DFT data for mixing enthalpy can take on very small values, we have instead reported the relative error with respect to the $2$-norm of the mixing enthalpy data, taken across all composition.
\FloatBarrier
While the relative errors for the mixing enthalpy tend to be large, the relative errors for the other physical properties are well below $0.2$. 

\section*{Data availability statement}
Additional data associated with this article are available upon request from the corresponding author (\texttt{ahegde@sandia.gov}).
\FloatBarrier
\bibliographystyle{ieeetr}
\bibliography{references}  

\end{document}

%% file: simulation_workflow.tex
\begin{tikzpicture}[node distance=1cm]
\node[draw, text width = 3.00cm, minimum height = 1cm, align = center] (IAP) {RAMPAGE \\ potential};
\node[draw] (E) [above = 1cm of IAP]  {Element A and B Info};
\node[draw, text width = 3.00cm,minimum height = 1cm, align = center, ] (LAMMPS) [right = 2cm of IAP] {LAMMPS};
\node[text width = 4.50cm, align = center] (other) [above =1.0cm of LAMMPS] { \begin{itemize} \setlength\itemsep{-2em} \item SQS (compositions) \\ \item numerical tolerances \\ \item etc. \end{itemize}};
\node[align = center] (theta) [left = 1cm of IAP] {\Large $\theta$};
\node[align = center] (y) [right = 1.5cm of LAMMPS] {\Large $f(\theta)$};
\draw[-{Latex}, thick] (E) -- (IAP.north);
\draw[-{Latex}, thick] (IAP) -- (LAMMPS);
\draw[-{Latex}, thick] (other) -- (LAMMPS);
\draw[-{Latex}, thick] (LAMMPS) -- (y);
\draw[-{Latex}, thick] (theta) -- (IAP);
\draw[thick,dotted]    ($(E.north west)+(-0.50,1.0)$) rectangle ($(LAMMPS.south east)+(0.7,-0.5)$) {};
\end{tikzpicture}

%% file: force_workflow.tex
\begin{tikzpicture}[node distance=1cm]
\node[draw, text width = 3.00cm, minimum height = 1cm, align = center] (IAP) {RAMPAGE \\ potential};
\node[draw, text width = 3.00cm, minimum height = 1cm, align = center, ] (LAMMPS) [right = 2cm of IAP] {LAMMPS};
\node[draw] (E) [above = 1cm of IAP]  {Element A and B Info};
\node[draw, text width = 2.00cm, minimum height = 0.5cm, align = center] (rescale) [above = 1.0cm of LAMMPS] {rescaling};
\node[draw, minimum width = 2cm, minimum height = 1cm, align = center] (DFT) [above = 1.0cm of rescale] {DFT-MD};
\node[align = center] (struct) [above right = 0.1cm and 0.1cm of rescale.north] {structure, time, \\ \& temperature.};
\node[align = center] (theta) [left = 1.0cm of IAP] {\huge $\theta$};
\node[align = center] (flmp) [right = 1.5cm of LAMMPS] {Force};
\node[align = center] (fdft) [right = 2.0cm of DFT] {Force};
\draw[-{Latex}, thick] (IAP) -- (LAMMPS);
\draw[-{Latex}, thick] (rescale) -- (LAMMPS);
\draw[-{Latex}, thick] (DFT) -- (rescale);
\draw[-{Latex}, thick] (LAMMPS) -- (flmp);
\draw[-{Latex}, thick] (DFT) -- (fdft);
\draw[-{Latex}, thick] (theta) -- (IAP);
\draw[-{Latex}, thick] (E) -- (IAP);
\end{tikzpicture}

%% file: main.bbl
\begin{thebibliography}{10}

\bibitem{brenner2000art}
D.~W. Brenner, ``The art and science of an analytic potential,'' {\em physica
  status solidi (b)}, vol.~217, no.~1, pp.~23--40, 2000.

\bibitem{goedecker1999linear}
S.~Goedecker, ``Linear scaling electronic structure methods,'' {\em Reviews of
  Modern Physics}, vol.~71, no.~4, p.~1085, 1999.

\bibitem{frederiksen2004bayesian}
S.~L. Frederiksen, K.~W. Jacobsen, K.~S. Brown, and J.~P. Sethna, ``Bayesian
  ensemble approach to error estimation of interatomic potentials,'' {\em
  Physical review letters}, vol.~93, no.~16, p.~165501, 2004.

\bibitem{angelikopoulos2012bayesian}
P.~Angelikopoulos, C.~Papadimitriou, and P.~Koumoutsakos, ``Bayesian
  uncertainty quantification and propagation in molecular dynamics simulations:
  a high performance computing framework,'' {\em The Journal of chemical
  physics}, vol.~137, no.~14, p.~144103, 2012.

\bibitem{dutta2018bayesian}
R.~Dutta, Z.~F. Brotzakis, and A.~Mira, ``Bayesian calibration of force-fields
  from experimental data: Tip4p water,'' {\em The Journal of chemical physics},
  vol.~149, no.~15, p.~154110, 2018.

\bibitem{longbottom2019uncertainty}
S.~Longbottom and P.~Brommer, ``Uncertainty quantification for classical
  effective potentials: an extension to potfit,'' {\em Modelling and Simulation
  in Materials Science and Engineering}, vol.~27, no.~4, p.~044001, 2019.

\bibitem{patrone2019uncertainty}
P.~N. Patrone and A.~Dienstfrey, ``Uncertainty quantification for molecular
  dynamics,'' {\em Reviews in Computational Chemistry}, vol.~31, pp.~115--169,
  2019.

\bibitem{cailliez2020bayesian}
F.~Cailliez, P.~Pernot, F.~Rizzi, R.~Jones, O.~Knio, G.~Arampatzis, and
  P.~Koumoutsakos, ``Bayesian calibration of force fields for molecular
  simulations,'' in {\em Uncertainty quantification in multiscale materials
  modeling} (Y.~Wang and D.~L. McDowell, eds.), ch.~6, pp.~169 -- 227, Woodhead
  Publishing, 2020.

\bibitem{vassaux2021ensembles}
M.~Vassaux, S.~Wan, W.~Edeling, and P.~V. Coveney, ``Ensembles are required to
  handle aleatoric and parametric uncertainty in molecular dynamics
  simulation,'' {\em Journal of chemical theory and computation}, vol.~17,
  no.~8, pp.~5187--5197, 2021.

\bibitem{cooke2008statistical}
B.~Cooke and S.~C. Schmidler, ``Statistical prediction and molecular dynamics
  simulation,'' {\em Biophysical journal}, vol.~95, no.~10, pp.~4497--4511,
  2008.

\bibitem{cailliez2011statistical}
F.~Cailliez and P.~Pernot, ``Statistical approaches to forcefield calibration
  and prediction uncertainty in molecular simulation,'' {\em The Journal of
  chemical physics}, vol.~134, no.~5, p.~054124, 2011.

\bibitem{pernot2017critical}
P.~Pernot and F.~Cailliez, ``A critical review of statistical
  calibration/prediction models handling data inconsistency and model
  inadequacy,'' {\em AIChE Journal}, vol.~63, no.~10, pp.~4642--4665, 2017.

\bibitem{rizzi2012uncertainty1}
F.~Rizzi, H.~N. Najm, B.~J. Debusschere, K.~Sargsyan, M.~Salloum,
  H.~Adalsteinsson, and O.~M. Knio, ``Uncertainty quantification in md
  simulations. part i: Forward propagation,'' {\em Multiscale Modeling \&
  Simulation}, vol.~10, no.~4, pp.~1428--1459, 2012.

\bibitem{rizzi2012uncertainty2}
F.~Rizzi, H.~N. Najm, B.~J. Debusschere, K.~Sargsyan, M.~Salloum,
  H.~Adalsteinsson, and O.~M. Knio, ``Uncertainty quantification in md
  simulations. part ii: Bayesian inference of force-field parameters,'' {\em
  Multiscale Modeling \& Simulation}, vol.~10, no.~4, pp.~1460--1492, 2012.

\bibitem{zhou2017uncertainty}
X.~Zhou and S.~M. Foiles, ``Uncertainty quantification and reduction of
  molecular dynamics models,'' in {\em Uncertainty Quantification and Model
  Calibration} (J.~P. Hessling, ed.), IntechOpen London, 2017.

\bibitem{chernatynskiy2013uncertainty}
A.~Chernatynskiy, S.~R. Phillpot, and R.~LeSar, ``Uncertainty quantification in
  multiscale simulation of materials: A prospective,'' {\em Annual Review of
  Materials Research}, vol.~43, pp.~157--182, 2013.

\bibitem{ward2012rapid}
L.~Ward, A.~Agrawal, K.~M. Flores, and W.~Windl, ``Rapid production of accurate
  embedded-atom method potentials for metal alloys,'' {\em arXiv},
  p.~1209.0619, 2012.

\bibitem{baskes1987application}
M.~Baskes, ``Application of the embedded-atom method to covalent materials: a
  semiempirical potential for silicon,'' {\em Physical review letters},
  vol.~59, no.~23, p.~2666, 1987.

\bibitem{baskes1992modified}
M.~I. Baskes, ``Modified embedded-atom potentials for cubic materials and
  impurities,'' {\em Physical review B}, vol.~46, no.~5, p.~2727, 1992.

\bibitem{thompson2015spectral}
A.~P. Thompson, L.~P. Swiler, C.~R. Trott, S.~M. Foiles, and G.~J. Tucker,
  ``Spectral neighbor analysis method for automated generation of
  quantum-accurate interatomic potentials,'' {\em Journal of Computational
  Physics}, vol.~285, pp.~316--330, 2015.

\bibitem{mueller2020machine}
T.~Mueller, A.~Hernandez, and C.~Wang, ``Machine learning for interatomic
  potential models,'' {\em The Journal of chemical physics}, vol.~152, no.~5,
  p.~050902, 2020.

\bibitem{becker2011atomistic}
C.~A. Becker, ``Atomistic simulations for engineering: Potentials and
  challenges,'' {\em Tools, models, databases and simulation tools developed
  and needed to realize the vision of ICME, ASM International, Materials Park},
  p.~91, 2011.

\bibitem{becker2013considerations}
C.~A. Becker, F.~Tavazza, Z.~T. Trautt, and R.~A.~B. de~Macedo,
  ``Considerations for choosing and using force fields and interatomic
  potentials in materials science and engineering,'' {\em Current Opinion in
  Solid State and Materials Science}, vol.~17, no.~6, pp.~277--283, 2013.

\bibitem{hale2018evaluating}
L.~M. Hale, Z.~T. Trautt, and C.~A. Becker, ``Evaluating variability with
  atomistic simulations: the effect of potential and calculation methodology on
  the modeling of lattice and elastic constants,'' {\em Modelling and
  Simulation in Materials Science and Engineering}, vol.~26, no.~5, p.~055003,
  2018.

\bibitem{tadmor2011potential}
E.~B. Tadmor, R.~S. Elliott, J.~P. Sethna, R.~E. Miller, and C.~A. Becker,
  ``The potential of atomistic simulations and the knowledgebase of interatomic
  models,'' {\em JOM}, vol.~63, no.~7, p.~17, 2011.

\bibitem{elliott2011knowledgebase}
R.~Elliott and E.~Tadmor, ``Knowledgebase of interatomic models (kim)
  application programming interface (api),'' {\em OpenKIM. URL https://openkim.
  org/kim-api}, 2011.

\bibitem{voter1994embedded}
A.~F. Voter, ``The embedded atom method,'' {\em Intermetallic Compounds:
  Principles and Practice}, vol.~1, p.~77, 1994.

\bibitem{grochola2005fitting}
G.~Grochola, S.~P. Russo, and I.~K. Snook, ``On fitting a gold embedded atom
  method potential using the force matching method,'' {\em The Journal of
  chemical physics}, vol.~123, no.~20, p.~204719, 2005.

\bibitem{MARTINEZ2013}
J.~A. Martinez, D.~E. Yilmaz, T.~Liang, S.~B. Sinnott, and S.~R. Phillpot,
  ``Fitting empirical potentials: Challenges and methodologies,'' {\em Current
  Opinion in Solid State and Materials Science}, vol.~17, no.~6, pp.~263--270,
  2013.
\newblock Frontiers in Methods for Materials Simulations.

\bibitem{gao2016high}
M.~C. Gao, J.-W. Yeh, P.~K. Liaw, and Y.~Zhang, {\em High-Entropy Alloys:
  Fundamentals and Applications}.
\newblock Springer International Publishing, 2016.

\bibitem{cantor2020multicomponent}
B.~Cantor, ``Multicomponent high-entropy cantor alloys,'' {\em Progress in
  Materials Science}, p.~100754, 2020.

\bibitem{becker_trautt_hale}
C.~A. Becker, Z.~T. Trautt, and L.~M. Hale, ``Overview.''

\bibitem{finnis1984simple}
M.~W. Finnis and J.~E. Sinclair, ``A simple empirical {N}-body potential for
  transition metals,'' {\em Philosophical Magazine A}, vol.~50, pp.~45--55,
  July 1984.

\bibitem{daw1984embedded}
M.~S. Daw and M.~I. Baskes, ``Embedded-atom method: {Derivation} and
  application to impurities, surfaces, and other defects in metals,'' {\em
  Physical Review B}, vol.~29, pp.~6443--6453, June 1984.

\bibitem{Agrawal_2013}
A.~Agrawal, R.~Mishra, L.~Ward, K.~M. Flores, and W.~Windl, ``An embedded atom
  method potential of beryllium,'' {\em Modelling and Simulation in Materials
  Science and Engineering}, vol.~21, p.~085001, oct 2013.

\bibitem{Agrawal_2015}
A.~Agrawal, R.~Mishra, L.~Ward, K.~M. Flores, and W.~Windl, ``Corrigendum: An
  embedded atom method potential of beryllium (modelling simul. mater. sci.
  eng. 21 085001),'' {\em Modelling and Simulation in Materials Science and
  Engineering}, vol.~23, p.~069501, aug 2015.

\bibitem{jin2015}
H.-S. Jin, J.-D. An, and Y.-S. Jong, ``Eam potentials for bcc, fcc and hcp
  metals with farther neighbor atoms,'' {\em Applied Physics A}, vol.~120, 07
  2015.

\bibitem{Roth2017}
P.~C. Roth, H.~Shan, D.~Riegner, N.~Antolin, S.~Sreepathi, L.~Oliker,
  S.~Williams, S.~Moore, and W.~Windl, ``Performance analysis and optimization
  of the rampage metal alloy potential generation software,'' in {\em
  Proceedings of the 4th ACM SIGPLAN International Workshop on Software
  Engineering for Parallel Systems}, SEPS 2017, (New York, NY, USA),
  p.~11–20, Association for Computing Machinery, 2017.

\bibitem{plimpton1995fast}
S.~Plimpton, ``Fast parallel algorithms for short-range molecular dynamics,''
  {\em Journal of computational physics}, vol.~117, no.~1, pp.~1--19, 1995.

\bibitem{voter1986accurate}
A.~F. Voter and S.~P. Chen, ``{Accurate interatomic potentials for Ni, Al and
  Ni$_3$Al},'' {\em MRS Online Proceedings Library (OPL)}, vol.~82, 1986.

\bibitem{riegnerThesis}
D.~Riegner, {\em Molecular Dynamics Simulations of Metallic Glass Formation and
  Structure}.
\newblock PhD thesis, Ohio State University, 2016.

\bibitem{weiss2022RAMPAGE}
E.~Weiss, D.~Riegner, and W.~Windl, ``Rapid production of accurate
  embedded-atom method potentials for metal alloys,'' {\em Manuscript in
  preparation}, 2022.

\bibitem{OBERDORFER2019}
C.~Oberdorfer and W.~Windl, ``Bond-order bond energy model for alloys,'' {\em
  Acta Materialia}, vol.~179, pp.~406--413, 2019.

\bibitem{von2010generation}
J.~von Pezold, A.~Dick, M.~Fri{\'a}k, and J.~Neugebauer, ``Generation and
  performance of special quasirandom structures for studying the elastic
  properties of random alloys: Application to al-ti,'' {\em Physical Review B},
  vol.~81, no.~9, p.~094203, 2010.

\bibitem{kresse1994ab}
G.~Kresse and J.~Hafner, ``Ab initio molecular-dynamics simulation of the
  liquid-metal--amorphous-semiconductor transition in germanium,'' {\em
  Physical Review B}, vol.~49, pp.~14251--14269, May 1994.

\bibitem{kresse1996efficiency}
G.~Kresse and J.~Furthmüller, ``Efficiency of ab-initio total energy
  calculations for metals and semiconductors using a plane-wave basis set,''
  {\em Computational Materials Science}, vol.~6, pp.~15--50, July 1996.

\bibitem{kresse1996efficient}
G.~Kresse and J.~Furthmüller, ``Efficient iterative schemes for ab initio
  total-energy calculations using a plane-wave basis set,'' {\em Physical
  Review B}, vol.~54, pp.~11169--11186, Oct. 1996.
\newblock Publisher: American Physical Society.

\bibitem{kresse1999ultrasoft}
G.~Kresse and D.~Joubert, ``From ultrasoft pseudopotentials to the projector
  augmented-wave method,'' {\em Physical Review B}, vol.~59, pp.~1758--1775,
  Jan. 1999.
\newblock Publisher: American Physical Society.

\bibitem{kennedy2001bayesian}
M.~C. Kennedy and A.~O'Hagan, ``Bayesian calibration of computer models,'' {\em
  Journal of the Royal Statistical Society: Series B (Statistical
  Methodology)}, vol.~63, no.~3, pp.~425--464, 2001.

\bibitem{higdon2004combining}
D.~Higdon, M.~Kennedy, J.~C. Cavendish, J.~A. Cafeo, and R.~D. Ryne,
  ``Combining field data and computer simulations for calibration and
  prediction,'' {\em SIAM Journal on Scientific Computing}, vol.~26, no.~2,
  pp.~448--466, 2004.

\bibitem{higdon2008computer}
D.~Higdon, J.~Gattiker, B.~Williams, and M.~Rightley, ``Computer model
  calibration using high-dimensional output,'' {\em Journal of the American
  Statistical Association}, vol.~103, no.~482, pp.~570--583, 2008.

\bibitem{bayarri2007framework}
M.~J. Bayarri, J.~O. Berger, R.~Paulo, J.~Sacks, J.~A. Cafeo, J.~Cavendish,
  C.-H. Lin, and J.~Tu, ``A framework for validation of computer models,'' {\em
  Technometrics}, vol.~49, no.~2, pp.~138--154, 2007.

\bibitem{sargsyan2015statistical}
K.~Sargsyan, H.~Najm, and R.~Ghanem, ``On the statistical calibration of
  physical models,'' {\em International Journal of Chemical Kinetics}, vol.~47,
  no.~4, pp.~246--276, 2015.

\bibitem{bernardo2009bayesian}
J.~M. Bernardo and A.~F. Smith, {\em Bayesian theory}, vol.~405.
\newblock John Wiley \& Sons, 2009.

\bibitem{gelman1995bayesian}
A.~Gelman, J.~B. Carlin, H.~S. Stern, and D.~B. Rubin, {\em Bayesian data
  analysis}.
\newblock Chapman and Hall/CRC, 1995.

\bibitem{plimpton2012computational}
S.~J. Plimpton and A.~P. Thompson, ``Computational aspects of many-body
  potentials,'' {\em MRS bulletin}, vol.~37, no.~5, pp.~513--521, 2012.

\bibitem{zuo2020performance}
Y.~Zuo, C.~Chen, X.~Li, Z.~Deng, Y.~Chen, J.~Behler, G.~Cs{\'a}nyi, A.~V.
  Shapeev, A.~P. Thompson, M.~A. Wood, and S.~P. Ong, ``Performance and cost
  assessment of machine learning interatomic potentials,'' {\em The Journal of
  Physical Chemistry A}, vol.~124, no.~4, pp.~731--745, 2020.

\bibitem{bishop2006pattern}
C.~M. Bishop, {\em Pattern recognition and machine learning}.
\newblock Springer, 2006.

\bibitem{mackay1996hyperparameters}
D.~J. MacKay, ``Hyperparameters: optimize, or integrate out?,'' in {\em Maximum
  entropy and bayesian methods}, pp.~43--59, Springer, 1996.

\bibitem{kass1995bayes}
R.~E. Kass and A.~E. Raftery, ``Bayes factors,'' {\em Journal of the american
  statistical association}, vol.~90, no.~430, pp.~773--795, 1995.

\bibitem{mackay1992bayesian}
D.~J. MacKay, ``Bayesian interpolation,'' {\em Neural computation}, vol.~4,
  no.~3, pp.~415--447, 1992.

\bibitem{gelman1996posterior}
A.~Gelman, X.-L. Meng, and H.~Stern, ``Posterior predictive assessment of model
  fitness via realized discrepancies,'' {\em Statistica sinica}, pp.~733--760,
  1996.

\bibitem{liu2008monte}
J.~S. Liu, {\em Monte Carlo strategies in scientific computing}.
\newblock Springer Science \& Business Media, 2008.

\bibitem{sacks1989design}
J.~Sacks, W.~J. Welch, T.~J. Mitchell, and H.~P. Wynn, ``Design and analysis of
  computer experiments,'' {\em Statistical science}, pp.~409--423, 1989.

\bibitem{stuart2018posterior}
A.~Stuart and A.~Teckentrup, ``{Posterior consistency for Gaussian process
  approximations of Bayesian posterior distributions},'' {\em Mathematics of
  Computation}, vol.~87, no.~310, pp.~721--753, 2018.

\bibitem{mckay1979comparison}
M.~D. McKay, R.~J. Beckman, and W.~J. Conover, ``A comparison of three methods
  for selecting values of input variables in the analysis of output from a
  computer code,'' {\em Technometrics}, vol.~21, no.~2, pp.~239--245, 1979.

\bibitem{zhou2004misfit}
X.~Zhou, R.~Johnson, and H.~Wadley, ``Misfit-energy-increasing dislocations in
  vapor-deposited cofe/nife multilayers,'' {\em Physical Review B}, vol.~69,
  no.~14, p.~144113, 2004.

\bibitem{voter1993alamos}
A.~F. Voter, ``Los alamos unclassified report la-ur 93-3901,'' 1993.

\bibitem{rasmussen2006gaussian}
C.~E. Rasmussen and C.~K. Williams, {\em Gaussian processes for machine
  learning}.
\newblock MIT Press, 2006.

\bibitem{brynjarsdottir2014learning}
J.~Brynjarsd{\' o}ttir and A.~O'Hagan, ``Learning about physical parameters:
  The importance of model discrepancy,'' {\em Inverse problems}, vol.~30,
  no.~11, p.~114007, 2014.

\bibitem{sargsyan2019embedded}
K.~Sargsyan, X.~Huan, and H.~N. Najm, ``Embedded model error representation for
  bayesian model calibration,'' {\em International Journal for Uncertainty
  Quantification}, vol.~9, no.~4, 2019.

\bibitem{wong2018optimizing}
Z.~M. Wong, T.~L. Tan, S.-W. Yang, and G.~Q. Xu, ``Optimizing special
  quasirandom structure (sqs) models for accurate functional property
  prediction in disordered 2d alloys,'' {\em Journal of Physics: Condensed
  Matter}, vol.~30, no.~48, p.~485402, 2018.

\bibitem{debusschere2017uqtk}
B.~Debusschere, K.~Sargsyan, C.~Safta, and K.~Chowdhary, ``The uncertainty
  quantification toolkit ({UQTk}),'' in {\em Handbook of Uncertainty
  Quantification} (R.~Ghanem, D.~Higdon, and H.~Owhadi, eds.), pp.~1807--1827,
  Springer, 2017.

\bibitem{debusschere2004numerical}
B.~J. Debusschere, H.~N. Najm, P.~P. P{\'e}bay, O.~M. Knio, R.~G. Ghanem, and
  O.~P. {Le Ma{\^\i}tre}, ``Numerical challenges in the use of polynomial chaos
  representations for stochastic processes,'' {\em SIAM journal on scientific
  computing}, vol.~26, no.~2, pp.~698--719, 2004.

\bibitem{hogg2018data}
D.~W. Hogg and D.~Foreman-Mackey, ``Data analysis recipes: Using markov chain
  monte carlo,'' {\em The Astrophysical Journal Supplement Series}, vol.~236,
  no.~1, p.~11, 2018.

\bibitem{Huan:2018b}
X.~Huan, C.~Safta, K.~Sargsyan, G.~Geraci, M.~S. Eldred, Z.~P. Vane, G.~Lacaze,
  J.~C. Oefelein, and H.~N. Najm, ``{Global Sensitivity Analysis and Estimation
  of Model Error, toward Uncertainty Quantification in Scramjet
  Computations},'' {\em AIAA Journal}, vol.~56, no.~3, pp.~1170--1184, 2018.

\bibitem{ercolessi1994interatomic}
F.~Ercolessi and J.~B. Adams, ``Interatomic potentials from first-principles
  calculations: the force-matching method,'' {\em EPL (Europhysics Letters)},
  vol.~26, no.~8, p.~583, 1994.

\bibitem{perdew1997generalized}
J.~P. Perdew, K.~Burke, and M.~Ernzerhof, ``Generalized {Gradient}
  {Approximation} {Made} {Simple},'' {\em Physical Review Letters}, vol.~78,
  pp.~1396--1396, Feb. 1997.

\bibitem{monkhorst1976special}
H.~J. Monkhorst and J.~D. Pack, ``Special points for {Brillouin}-zone
  integrations,'' {\em Physical Review B}, vol.~13, pp.~5188--5192, June 1976.

\bibitem{VASPKIT}
V.~Wang, N.~Xu, J.-C. Liu, G.~Tang, and W.-T. Geng, ``Vaspkit: A user-friendly
  interface facilitating high-throughput computing and analysis using vasp
  code,'' {\em Computer Physics Communications}, vol.~267, p.~108033, 2021.

\bibitem{vandewalle2013efficient}
A.~van~de Walle, P.~Tiwary, M.~De~Jong, D.~Olmsted, M.~Asta, A.~Dick, D.~Shin,
  Y.~Wang, L.-Q. Chen, and Z.-K. Liu, ``Efficient stochastic generation of
  special quasirandom structures,'' {\em CALPHAD (Computer Coupling of Phase
  Diagrams and Thermochemistry)}, vol.~42, 1 2013.

\end{thebibliography}
